\documentclass[conference, dvipsnames]{IEEEtran}

\usepackage{amsmath}
\usepackage{amssymb}
\usepackage{graphicx}
\usepackage{subfigure}
\usepackage{cite}
\usepackage{color}
\usepackage{accents}
\usepackage{url}
\usepackage{lipsum}
\usepackage[nolist]{acronym}

\usepackage{color}

\usepackage{pgfplots}
\pgfplotsset{compat = 1.15,
	yticklabel style={/pgf/number format/fixed}
}

\usepackage{tikz}
\usetikzlibrary{shapes}
\usetikzlibrary{shapes.geometric}
\usetikzlibrary{shapes.misc}
\usetikzlibrary{arrows.meta,arrows}
\usetikzlibrary{graphs}
\usetikzlibrary{fit}
\usetikzlibrary{positioning}
\usetikzlibrary{calc}
\usetikzlibrary{backgrounds}
\usetikzlibrary{decorations.pathreplacing}

\def \scalePicture{.7}
\def \scaleFigure{.5}

\newcommand{\tSNR}{\text{SNR}}

\newcommand{\diag}{\mathbf{diag}}

\newcommand{\tcr}{\textcolor{red}}
\newcommand{\tbl}{\textcolor{blue}}
\newcommand{\tpl}{\textcolor{Plum}}

\IEEEoverridecommandlockouts

\begin{acronym}
	\acro{AWGN}[AWGN]{additive white Gaussian noise}
	\acro{MIMO}[MIMO]{multiple-input and multiple-output}
	\acro{OFDM}[OFDM]{orthogonal frequency-division multiplexing}
	\acro{SVM}[SVM]{support vector machine}
	\acro{MMSE}[MMSE]{minimum mean-square error}
	\acro{ADC}[ADC]{analog-to-digital converter}
	\acro{iid}[i.i.d.]{independent and identically distributed}
\end{acronym}

\begin{document}
\title{Joint Channel Estimation and Data Decoding using SVM-based Receivers}

\author{Sami Ak{\i}n, Maxim Penner, and J\"{u}rgen Peissig\\
	Institute of Communications Technology\\
	Leibniz Universit\"{a}t Hannover\\
	Email: \{sami.akin, maxim.penner, and peissig\}@ikt.uni-hannover.de
	\thanks{This work was supported by the German Research Foundation (DFG) -- FeelMaTyC (329885056)}
}

\maketitle

\begin{abstract}
Modern communication systems organize receivers in blocks in order to simplify their analysis and design. However, an approach that considers the receiver design from a wider perspective rather than treating it block-by-block may take advantage of the impacts of these blocks on each other and provide better performance. Herein, we can benefit from machine learning and compose a receiver model implementing supervised learning techniques. With this motivation, we consider a one-to-one transmission system over a flat fast fading wireless channel and propose a \ac{SVM}-based receiver that combines the pilot-based channel estimation, data demodulation and decoding processes in one joint operation. We follow two techniques in the receiver design. We first design one \ac{SVM}-based classifier that outputs the class of the encoding codeword that enters the encoder at the transmitter side. Then, we put forward a model with one \ac{SVM}-based classifier per one bit in the encoding codeword, where each classifier assigns the value of the corresponding bit in the encoding vector. With the second technique, we simplify the receiver design especially for longer encoding codewords. We show that the \ac{SVM}-based receiver performs very closely to the maximum likelihood decoder, which is known to be the optimal decoding strategy when the encoding vectors at the transmitter are equally likely. We further show that the \ac{SVM}-based receiver outperforms the conventional receivers that perform channel estimation, data demodulation and decoding in blocks. Finally, we show that we can train the \ac{SVM}-based receiver with 1-bit \ac{ADC} outputs and the \ac{SVM}-based receiver can perform very closely to the conventional receivers that take 32-bit \ac{ADC} outputs as inputs. 
\end{abstract}

\acresetall
\section{Introduction}\label{sec:Introduction}
The design process in  modern communication systems usually involves dividing a transmission chain into several blocks and describing each block independently by invoking mathematical models. This applies to channel coding/decoding, data modulation/demodulation, modeling and estimating wireless channel gains, and hardware imperfections. Although it may be advantageous to investigate these blocks jointly, it is difficult to understand the generally non-linear relationships between the blocks and even more challenging to exploit these relationships. Nevertheless, following the advances in classification and clustering techniques, for example \cite{rawat2017deep,krizhevsky2012imagenet}, there have recently been efforts to transfer the success of machine learning to wireless communications \cite{alsheikh2014machine, jiang2016machine, lv2017machine, zhang2019deep}. Therefore, machine learning can make it possible in future to design an entire communication system on the basis of large training data-sets, for instance, obtained from wireless channel measurements, or at least, machine learning may become a supplement for existing mathematical models that do not lead to closed-form solutions.

In this regard, the authors in \cite{o2017introduction} explored several new fundamental opportunities offered by machine learning, in particular deep learning, in the physical layer. They trained one neural network per transmitter and receiver, separated by an additive white Gaussian noise channel, in order to adopt the channel conditions, and achieved the same transmission performance observed in handcrafted system designs and even outperformed them by learning joint modulation and coding schemes in certain scenarios. In \cite{jiang2016machine}, the authors provided an overview of where different machine learning algorithms could be applied in the fifth generation (5G) cellular network technology. Some application areas among many are \ac{MIMO} channel training, cognitive spectrum sensing and energy harvesting. Furthermore, the authors in \cite{ye2017power} used deep learning tools for channel estimation and signal detection in \ac{OFDM} systems.

Among the classical machine learning algorithms, \acp{SVM} \cite{Cortes:1995:SN:218919.218929} have become one of the most popular methods. The main advantages of \acp{SVM} are their ability to learn the non-linear separations of many training sets by choosing appropriate kernels and convex loss functions with a guaranteed global maximum \cite{1640771}. In a wireless communication system, an \ac{SVM}-based component can be deployed in several processes, e.g., channel estimation and data decoding. For instance, the authors in \cite{1642708} designed an \ac{SVM}-based channel estimator for \ac{OFDM} systems and compared it with a classical least-squares estimator. They showed that both of the estimators perform the same in high signal-to-noise regimes but the \ac{SVM}-based estimator outperforms the other in low signal-to-noise ratio regimes. Likewise, the authors in \cite{8285585} compared an \ac{SVM}-based channel estimator with least-squares and neural-network based estimators in Long-Term Evolution systems considering non-linear regression and showed that the SVM-based estimator performs better than the other estimators. Herein, we further refer interested readers to the other \ac{SVM}-based channel estimation techniques used in \ac{MIMO} channels \cite{charrada:tel-01350731,sanchez2004svm}.

In this paper, we consider a one-to-one transmission system over a flat fast fading wireless channel. Different from the aforementioned studies, we propose a machine learning-based, specifically an \ac{SVM}-based, receiver that combines the pilot-based channel estimation, data demodulation and decoding processes in one joint operation. Specifically, the receiver takes the pilot and data symbol outputs of the channel, i.e., the IQ-samples of the transmitted pilot and data symbols, and provides the encoding vector, i.e., the data vector entering the channel encoder at the transmitter side, in one process.

We follow two techniques in the receiver design. In the first design, we utilize one SVM-based classifier that outputs the class of the encoding codeword. In the second approach, we have as many SVM-based binary classifiers as the bits in one encoding vector, where each SVM-based classifier returns the value of the corresponding bit in the encoding vector. The second technique significantly reduces the classification complexity especially for longer encoding codewords. We train the SVM-based classifiers in channels that are modeled as first-order Gauss-Markov processes. Then, we compare the SVM-based receiver with two conventional receiver models, i.e., the receiver employs \ac{MMSE} and iterative Kalman filter estimators to perform channel estimation, respectively, and then demodulates and decodes the encoding codeword. The SVM-based receiver outperforms these two receivers. We further show that the SVM-based receiver can perform very closely to the maximum likelihood decoder, which is known to be an optimal decoding strategy when the encoding vectors are equally likely. Finally, we train the SVM-based receivers using 1-bit \ac{ADC} outputs and show that they perform very closely to the conventional receiver models that use 32-bit \ac{ADC} outputs.

\section{System Model}\label{sec:system_model}
We consider a system where a transmitter and a receiver communicate over a wireless channel. In the sequel, we initially describe the transmitter and the input-output relation in the wireless channel. Then, we discuss the ideal receiver model, two models in practice, and our machine learning based models, sequentially.
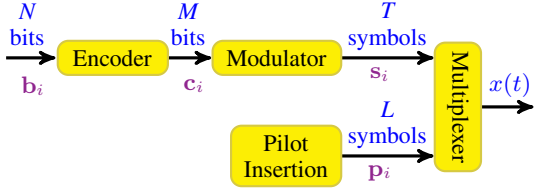
\begin{figure}
    \begin{center}
        \begin{tikzpicture}[
>=stealth',
black!50,
line width=0.7mm*\scalePicture,
text=black,
font = \small,
every new ->/.style = {shorten >=1pt*\scalePicture},
graphs/every graph/.style = {edges=rounded corners},
block/.style ={font = \small, rectangle, draw=blue!60!black, thick, fill = white, text width=3em*\scalePicture, align=center, rounded corners, minimum height=3.6em*\scalePicture},
point/.style={rectangle, minimum height=1.0em*\scalePicture, text width=3em*\scalePicture, draw = white, fill = white}]

    \matrix[column sep=12mm*\scalePicture, row sep=2mm*\scalePicture]
    {
        \node [point] (p11) {}; & \node [point] (p12) {}; & \node [point] (p13) {}; & \node [point] (p14) {};\\
        \node [point] (p21) {}; & \node [point] (p22) {}; & \node [point] (p23) {}; & \node [point] (p24) {};\\
        \node [point] (p31) {}; & \node [point] (p32) {}; & \node [point] (p33) {}; & \node [point] (p34) {};\\
        \node [point] (p41) {}; & \node [point] (p42) {}; & \node [point] (p43) {}; & \node [point] (p44) {};\\
    };


    \coordinate (t11) at ($(p11)$) {};
    \coordinate (t12) at ($(p12) + (180:49pt*\scalePicture)$) {};
    \draw [black, solid, line width=0.6mm*\scalePicture] (t11) [->] -- (t12);

    \coordinate (t21) at ($(p12)$) {};
    \coordinate (t22) at ($(p13) + (180:40.5pt*\scalePicture)$) {};
    \draw [black, solid, line width=0.6mm*\scalePicture] (t21) [->] -- (t22);

    \coordinate (t31) at ($(p13)$) {};
    \coordinate (t32) at ($(p14) + (0:4pt*\scalePicture)$) {};
    \draw [black, solid, line width=0.6mm*\scalePicture] (t31) [->] -- (t32);

    \coordinate (t41) at ($(p43)$) {};
    \coordinate (t42) at ($(p44) + (0:4pt*\scalePicture)$) {};
    \draw [black, solid, line width=0.6mm*\scalePicture] (t41) [->] -- (t42);

    \coordinate (t51) at ($(p24)!0.5!(p34) + (0:8pt*\scalePicture)$) {};
    \coordinate (t52) at ($(p24)!0.5!(p34) + (0:58pt*\scalePicture)$) {};
    \draw [black, solid, line width=0.6mm*\scalePicture] (t51) [->] -- (t52);


    \node [block, draw = yellow!80!black, fill = yellow!99!black, text width=5em*\scalePicture, minimum height=1.2em*\scalePicture, rotate=0] (node1) at ($(p12)+(180:18pt*\scalePicture)$) {Encoder};

    \node [block, draw = yellow!80!black, fill = yellow!99!black, text width=6em*\scalePicture, minimum height=1.2em*\scalePicture, rotate=0] (node2) at ($(p13)+(180:5pt*\scalePicture)$) {Modulator};

    \node [block, draw = yellow!80!black, fill = yellow!99!black, text width=5em*\scalePicture, minimum height=2.5em*\scalePicture, rotate=0] (node3) at ($(p43)$) {Pilot\\Insertion};

    \coordinate (cMult) at ($(p24)!0.5!(p34) + (0:12pt)$);
    \node [block, draw = yellow!80!black, fill = yellow!99!black, text width=6em*\scalePicture, minimum height=2.5em*\scalePicture, rotate=270] (node4) at ($(cMult)$) {Multiplexer};


    \coordinate (text11) at ($(p11)!0.15!(p12) + (90:18pt*\scalePicture)$);
    \coordinate (text12) at ($(p11)!0.15!(p12) + (270:15pt*\scalePicture)$);
    \node [rotate = 0, text width=2em*\scalePicture, align=center] (nodetext11) at (text11) {\tbl{\textit{N} bits}};
    \node [rotate = 0, text width=3em*\scalePicture, align=center] (nodetext12) at (text12) {$\hspace{0.5em} \tpl{\mathbf{b}_{i}}$};

    \coordinate (text21) at ($(p12)!0.3!(p13) + (90:18pt*\scalePicture)$);
    \coordinate (text22) at ($(p12)!0.3!(p13) + (270:15pt*\scalePicture)$);
    \node [rotate = 0, text width=2em*\scalePicture, align=center] (nodetext21) at (text21) {\tbl{\textit{M} bits}};
    \node [rotate = 0, text width=4em*\scalePicture, align=center] (nodetext22) at (text22) {$\hspace{0.5em} \tpl{\mathbf{c}_{i}}$};

    \coordinate (text31) at ($(p13)!0.5!(p14) + (0:17pt*\scalePicture) + (90:17pt*\scalePicture)$);
    \coordinate (text32) at ($(p13)!0.5!(p14) + (270:9pt*\scalePicture)$);
    \node [rotate = 0, text width=5em*\scalePicture, align=center] (nodetext31) at (text31) {\tbl{\textit{T} \textrm{symbols}}};
    \node [rotate = 0, text width=4em*\scalePicture, align=center] (nodetext32) at (text32) {$\hspace{2em} \tpl{\mathbf{s}_{i}}$};

    \coordinate (text41) at ($(p43)!0.5!(p44) + (0:17pt*\scalePicture) + (90:17pt*\scalePicture)$);
    \coordinate (text42) at ($(p43)!0.5!(p44) + (270:9pt*\scalePicture)$);
    \node [rotate = 0, text width=5em*\scalePicture, align=center] (nodetext41) at (text41) {\tbl{\textit{L} \textrm{symbols}}};
    \node [rotate = 0, text width=4em*\scalePicture, align=center] (nodetext42) at (text42) {$\hspace{2em} \tpl{\mathbf{p}_{i}}$};

    \coordinate (text52) at ($(t51)!0.75!(t52) + (90:12pt*\scalePicture)$);
    \node [rotate = 0, text width=4em*\scalePicture, align=center] (nodetext52) at (text52) {$\tbl{x(t)}$};

\end{tikzpicture}
        \caption{Transmitter. $L$ pilot symbols are inserted before every \emph{T} data symbols.}\label{fig:fig_1}
    \end{center}
	\vspace{-0.5cm}
\end{figure}
~
\subsubsection{Transmitter}\label{sec:transmitter}
The transmitter encodes a vector of $N$ bits, $\textbf{b}_{i}$, into a vector of $M$ bits, $\textbf{c}_{i}$, and then modulates $\textbf{c}_{i}$ into a vector of $T$ symbols, $\textbf{s}_{i}$, as seen in Fig. \ref{fig:fig_1}. Each encoding vector, $\textbf{b}_{i}$, is the $N\times1$ binary representation of one number from set $\mathcal{B}=\{0,1,\cdots,2^{N}-1\}$, and each number is chosen equally likely, i.e., $\Pr\{g(\textbf{b}_{i}) = B\} = \frac{1}{2^{N}}$ for $B\in\mathcal{B}$, where $g(\cdot)$ is the conversion from binary to decimal. Eventually multiplexing $\textbf{s}_{i}$ with a vector of $L$ pilot symbols, $\textbf{p}_{i}$, the transmitter sends the data to the receiver over the wireless channel. Here, the transmitter sends $\textbf{p}_{i}$ before $\textbf{s}_{i}$. The receiver knows the pilot symbols and when the transmitter sends them. The sub-index $i\in\{0,1,\cdots,\}$ is the frame index and each frame consists of $L+T$ symbols. Particularly, the transmitter sends a vector of pilot symbols every $L+T$ symbols, and $T$ data symbols after each pilot symbol vector. Here, we follow the periodic pilot placement model in \cite{dong2003optimal}, and impose the following average power constraint: $\mathbb{E}\left\{||\textbf{p}_{i}||^{2}\right\} + \mathbb{E}\left\{||\textbf{s}_{i}||^{2}\right\} \leq (L+T)P$, where $\mathbb{E}\{\cdot\}$ is the expectation operator. The total average transmission power allocated to $L$ pilot and $T$ data symbols is limited by $(L+T)P$ over a duration of $L+T$ symbols\footnote{For the sake of clarity, we place each pilot symbol vector before one data vector. However, one can insert several pilot symbol vectors in between the symbols of a data vector periodically, and re-express the power constraint accordingly. In addition, we employ only non-zero pilot symbols because we consider a flat fading channel. However, one can embed zero-valued symbols at beginning and end of a pilot symbol vector, as described in \cite{dong2003optimal}, if the channel is multi-path fading.}.
\begin{figure}
    \begin{center}
        \begin{tikzpicture}[
>=stealth',
black!50,
line width=0.7mm*\scalePicture,
text=black,
font = \small,
every new ->/.style = {shorten >=1pt*\scalePicture},
graphs/every graph/.style = {edges=rounded corners},
block/.style ={font = \small, rectangle, draw=blue!60!black, thick, fill = white, text width=2em*\scalePicture, align=center, rounded corners, minimum height=3.em*\scalePicture},
point/.style={rectangle, minimum height=0.2em*\scalePicture, text width=3em*\scalePicture, draw = white, fill = white},
cross/.style={cross out, draw=black, minimum size=1.0em*\scalePicture), inner sep=3pt*\scalePicture, outer sep=2pt*\scalePicture},
cross/.default={1pt*\scalePicture}
]

    \matrix[column sep=7mm*\scalePicture, row sep=1mm*\scalePicture]
    {
        \node [point] (p12) {}; & \node [point] (p13) {};\\
        \node [point] (p22) {}; & \node [point] (p23) {};\\
    };
    
    
    \coordinate (t11) at ($(p12) + (180:45pt*\scalePicture)$) {};
    \coordinate (t12) at ($(p12) + (180:14pt*\scalePicture)$) {};
    \draw [black, solid, line width=0.6mm*\scalePicture] (t11) [->] -- (t12);
    
    \coordinate (t121) at ($(p12) + (270:14pt*\scalePicture)$) {};
    \coordinate (t122) at ($(p22) + (270:22pt*\scalePicture)$) {};
    \draw [black, solid, line width=0.6mm*\scalePicture] (t122) [->] -- (t121);
    
    \coordinate (t21) at ($(p12)$) {};
    \coordinate (t22) at ($(p13) + (180:14pt*\scalePicture)$) {};
    \draw [black, solid, line width=0.6mm*\scalePicture] (t21) [->] -- (t22);
    
    \coordinate (t221) at ($(p13) + (270:14pt*\scalePicture)$) {};
    \coordinate (t222) at ($(p23) + (270:22pt*\scalePicture)$) {};
    \draw [black, solid, line width=0.6mm*\scalePicture] (t222) [->] -- (t221);
    
    \coordinate (t31) at ($(p13)$) {};
    \coordinate (t32) at ($(p13) + (0:45pt*\scalePicture)$) {};
    \draw [black, solid, line width=0.6mm*\scalePicture] (t31) [->] -- (t32);
    
    
    \node [circle, draw = blue!80!black, fill = ProcessBlue!50!white, text width=1.5em*\scalePicture, align = center] (fading) at ($(p12)$) {};
    \draw ($(p12)$) node[cross,red] {};
    
    \node [circle, draw = blue!80!black, fill = ProcessBlue!50!white, text width=1.5em*\scalePicture, align = center] (fading) at ($(p13)$) {};
    \draw ($(p13)$) node[cross, red, rotate = 45] {};

    
    \coordinate (text11) at ($(t11)!0.45!(t12) + (90:9pt*\scalePicture)$);
    \node [rotate = 0, text width=3em*\scalePicture, align=center] (nodetext1) at (text11) {$\tbl{x(t)}$};
    
    \coordinate (text12) at ($(t122)!0.5!(t121) + (0:18pt*\scalePicture)$);
    \node [rotate = 0, text width=3em*\scalePicture, align=center] (nodetext2) at (text12) {$\tcr{h(t)}$};
    
    \coordinate (text13) at ($(t122) + (270:18pt*\scalePicture)$);
    \node [rotate = 0, text width=5em*\scalePicture, align=center] (nodetext3) at (text13) {\tcr{\textrm{Channel\\Fading}}};
    
    \coordinate (text14) at ($(t222)!0.5!(t221) + (0:18pt*\scalePicture)$);
    \node [rotate = 0, text width=4em*\scalePicture, align=center] (nodetext4) at (text14) {$\tcr{w(t)}$};
    
    \coordinate (text15) at ($(t222) + (270:18pt*\scalePicture)$);
    \node [rotate = 0, text width=4em*\scalePicture, align=center] (nodetext5) at (text15) {\tcr{\textrm{Noise}}};
    
    \coordinate (text16) at ($(t31)!0.55!(t32) + (90:9pt*\scalePicture)$);
    \node [rotate = 0, text width=4em*\scalePicture, align=center] (nodetext1) at (text16) {$\color{orange}{y(t)}$};

\end{tikzpicture}

%
%
%
%
%
%
%
        \caption{Wireless channel.}\label{fig:fig_2}
    \end{center}
	\vspace{-0.5cm}
\end{figure}
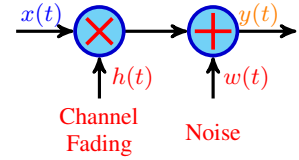
~
\subsubsection{Channel input-output}\label{sec:channel_input_output}
We describe the discrete channel input-output relation at time instant $t$ during transmission as
\begin{equation}\label{eq:channel_input_output}
y(t) = h(t)x(t) + w(t)\text{ for }t\in\{0,1,\cdots\},
\end{equation}
as seen in Fig. \ref{fig:fig_2}. Above, $x(t)$ and $y(t)$ are the channel input and output, respectively, and $h(t)$ denotes the channel fading gain and can have an arbitrary marginal distribution with finite power, i.e., $\mathbb{E}\left\{|h(t)|^{2}\right\} = \sigma_{h}^{2}<\infty$. We assume a flat fading channel and that the fading gains are correlated over time. We model the fading process in the channel as a first-order Gauss-Markov process and describe it as
\begin{equation}\label{eq:channel_model}
h(t) = \alpha h(t-1) + \beta(t),
\end{equation}
where $0\leq\alpha\leq1$, and $\{\beta(t)\}$'s are \ac{iid}, zero-mean and circularly-symmetric complex Gaussian variables with variance equal to $(1-\alpha^2)\sigma_{h}^{2}$. In the above formulation, we have $\alpha$ as the parameter that controls the rate of the channel variations between consecutive transmissions. For example, when $\alpha = 0$, the fading coefficients are uncorrelated, whereas when $\alpha = 1$ the fading coefficients remain constant over the transmission duration. For bandwidths in the $10$ kHz range and Doppler spreads of the order of $100$ Hz, $\alpha$ will typically range between $0.9$ and $0.99$ \cite{abou2005binary}.

In (\ref{eq:channel_input_output}), $w(t)$ models the additive thermal noise, and is a zero-mean and circularly-symmetric complex Gaussian random variable with variance $\mathbb{E}\left\{|w(t)|^{2}\right\} = \sigma_{w}^{2}$. We further assume that $\{w(t)\}$ is an \ac{iid} sequence. Herein, regarding a unit sampling time, we assume that the transmitter sends the first pilot symbol of the $i^{\text{th}}$ frame at time instant $t = i(L+T)$, and hence the data symbols of the same frame at time instants between $t=i(L+T)+L$ and $t = (i+1)(L+T)-1$. Particularly, we have $x(i(L+T)+j) = p_{i}(j)$ for $j\in\{0,\cdots,L-1\}$, and $x(i(L+T)+ L+j) = s_{i}(j)$ for $j\in\{0,\cdots,T-1\}$. Furthermore, we define the pilot vector transmitted before the $i^{\text{th}}$ data vector as $\textbf{p}_{i} = [p^{\dagger}_{i}(L-1), \cdots, p^{\dagger}_{i}(0)]^{\dagger}$ and the data vector as $\textbf{s}_{i} = [s^{\dagger}_{i}(T-1),\cdots,s^{\dagger}_{i}(0)]^{\dagger}$. Above, $\dagger$ denotes the conjugate transpose operator. Here, we follow a vector notation where the last transmitted symbol is the first element of the vector for the sake convenience in the presented analysis throughout the paper.

\begin{figure}
	\begin{center}
		\begin{tikzpicture}[
    >=stealth',
    black!50,
    line width=0.7mm*\scalePicture,
    text=black,
    font = \small,
    every new ->/.style = {shorten >=1pt*\scalePicture},
    graphs/every graph/.style = {edges=rounded corners},
    block/.style ={font = \small, rectangle, draw=blue!60!black, thick, fill = white, text width=3em*\scalePicture, align=center, rounded corners, minimum height=3.6em*\scalePicture},
    point/.style={rectangle, minimum height=1.0em*\scalePicture, text width=3em*\scalePicture, draw = white, fill = white},
    cross/.style={cross out, draw=black, minimum size=2.0em*\scalePicture), inner sep=3pt*\scalePicture, outer sep=2pt*\scalePicture},
    cross/.default={1pt*\scalePicture}
    ]
    \matrix[column sep=7mm*\scalePicture, row sep=5mm*\scalePicture]
    {
        \node [point] (p11) {}; & \node [point] (p12) {}; & \node [point] (p13) {}; & \node [point] (p14) {};\\
        \node [point] (p21) {}; & \node [point] (p22) {}; & \node [point] (p23) {}; & \node [point] (p24) {};\\
        \node [point] (p31) {}; & \node [point] (p32) {}; & \node [point] (p33) {}; & \node [point] (p34) {};\\
        \node [point] (p41) {}; & \node [point] (p42) {}; & \node [point] (p43) {}; & \node [point] (p44) {};\\
    };
    
    
    \coordinate (t11) at ($(p21)!0.5!(p31) + (180:44.5pt*\scalePicture)$) {};
    \coordinate (t12) at ($(p21)!0.5!(p31) + (180:13.5pt*\scalePicture)$) {};
    \draw [black, solid, line width=0.6mm*\scalePicture] (t11) [->] -- (t12);

    \coordinate (t21) at ($(p21) + (0:10pt*\scalePicture) + (90:10pt*\scalePicture)$) {};
    \coordinate (t22) at ($(p22) + (180:1.5pt*\scalePicture) + (90:10pt*\scalePicture)$) {};
    \draw [black, solid, line width=0.6mm*\scalePicture] (t21) [->] -- (t22);
    
    \coordinate (t31) at ($(p31) + (0:10pt*\scalePicture) + (270:10pt*\scalePicture)$) {};
    \coordinate (t32) at ($(p32) + (180:1.5pt*\scalePicture) + (270:10pt*\scalePicture)$) {};
    \draw [black, solid, line width=0.6mm*\scalePicture] (t31) [->] -- (t32);
    
    \coordinate (t41) at ($(p22)!0.5!(p32) + (0:150pt*\scalePicture)$) {};
    \coordinate (t42) at ($(p22)!0.5!(p32) + (0:200pt*\scalePicture)$) {};
    \draw [black, solid, line width=0.6mm*\scalePicture] (t41) [->] -- (t42);
    
    
    \coordinate (Demultiplexer) at ($(p21)!0.5!(p31)$) {};
    \node [block, draw = black!20!Cyan, fill = white!60!Cyan, text width=9em*\scalePicture, minimum height=2.5em*\scalePicture, rotate=270] (node1) at ($(Demultiplexer)$) {De-multiplexer};
    
    \coordinate (Receiver) at ($(p22)!0.5!(p32) + (0:79pt*\scalePicture)$) {};
    \node [block, draw = black!20!Cyan, fill = white!60!Cyan, text width=15em*\scalePicture, minimum height=2.0em*\scalePicture, rotate=0] (node2) at ($(Receiver)$) {Ideal Receiver:\\\text{ }\\$\tpl{\mathbf{\widehat{b}}_{i}}$ \color{black}{$= \arg\max$} $\Big\{\tpl{\mathbf{b}_{i}}:\Pr\{\tpl{\mathbf{b}_{i}}|\tpl{\mathbf{z}_{i}},\tpl{\mathbf{r}_{i}},\tpl{\mathbf{r}_{i+1}}\}\Big\}$};
    
    
    \coordinate (text1) at ($(t11)!0.4!(t12) + (90:9pt*\scalePicture)$) {};
    \node [rotate = 0, text width=4em*\scalePicture, align=center] (nodetext1) at (text1) {$\color{orange}{y(t)}$};
    
    \coordinate (RDS) at ($(t21)!0.5!(t22) + (0:7pt*\scalePicture) + (90:32pt*\scalePicture)$) {};
    \node [rotate = 0, text width=12em*\scalePicture, align=center] (nodetextRDS) at (RDS) {\color{orange}{Received Data\\Symbols}};
    
    \coordinate (text2) at ($(t21)!0.5!(t22) + (270:12pt*\scalePicture)$) {};
    \node [rotate = 0, text width=4em*\scalePicture, align=center] (nodetext2) at (text2) {$\tpl{\mathbf{z}_{i}}$};
    
    \coordinate (text3) at ($(t31)!0.5!(t32) + (90:12pt*\scalePicture)$) {};
    \node [rotate = 0, text width=4em*\scalePicture, align=center] (nodetext3) at (text3) {$\tpl{\mathbf{r}_{i+1}}, \tpl{\mathbf{r}_{i}}$};
    
    \coordinate (RPS) at ($(t31)!0.5!(t32) + (0:7pt*\scalePicture) + (270:32pt*\scalePicture)$) {};
    \node [rotate = 0, text width=12em*\scalePicture, align=center] (nodetextRPS) at (RPS) {\color{orange}{Received\\Pilot Symbols}};
    
    \coordinate (text4) at ($(t41)!0.45!(t42) + (90:18pt*\scalePicture)$) {};
    \node [rotate = 0, text width=2em*\scalePicture, align=center] (nodetext4) at (text4) {\color{orange}{$N$ bits}};
    
    \coordinate (text5) at ($(t41)!0.5!(t42) + (270:12pt*\scalePicture)$) {};
    \node [rotate = 0, text width=4em*\scalePicture, align=center] (nodetext5) at (text5) {$\tpl{\mathbf{\widehat{b}}_{i}}$};
    
\end{tikzpicture}
		\caption{Ideal receiver.}\label{fig:fig_3}
	\end{center}
	\vspace{-0.5cm}
\end{figure}
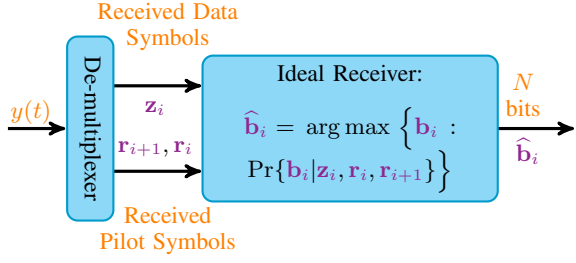
\subsubsection{Receiver}\label{sec:receiver}
The channel output, $y(t)$, following the de-multiplexer, becomes $r_{i}(j) = y(i(L+T)+j)$ and $z_{i}(j) = y(i(L+T)+L+j)$ when the transmitter sends $p_{i}(j)$ and $s_{i}(j)$, respectively. We denote the received $i^{\text{th}}$ pilot and data vectors by $\textbf{r}_{i} = [r^{\dagger}_{i}(L-1), \cdots, r^{\dagger}_{i}(0)]^{\dagger}$ and $\textbf{z}_{i} = [z^{\dagger}_{i}(T-1),\cdots,z^{\dagger}_{i}(0)]^{\dagger}$, respectively. The receiver uses the received $i^{\text{th}}$ data vector, $\textbf{z}_{i}$, and the pilot symbol vectors transmitted before and after $\textbf{z}_{i}$, i.e., $\textbf{r}_{i}$ and $\textbf{r}_{i+1}$, in order to decode the transmitted encoding vector, $\textbf{b}_{i}$. We consider only $\textbf{r}_{i}$ and $\textbf{r}_{i+1}$ in the decoding process because the channel fading gain, $h(t)$, changes over time as seen in (\ref{eq:channel_model}), and the correlation between the fading gains decreases by time. Therefore, it is enough to include the pilot symbols around the data vector since the pilot symbols far from the data vector will not bring significant improvement in the decoding process. Here, we also assume that the receiver knows the second order channel statistics, i.e., the flat fading and $\alpha$.

\paragraph{Ideal Receiver}\label{para:idealReceiver} As seen in Fig. \ref{fig:fig_3}, the ideal receiver is the one that chooses a number from the encoding set, $\mathcal{B}$, such that the probability $\Pr\{\textbf{b}_{i}|\textbf{z}_{i}, \textbf{r}_{i}, \textbf{r}_{i+1}\}$ is maximized. Recall that we define $g(\textbf{b}_{i}) = B$ as the conversion from binary to decimal and $B\in\mathcal{B}=\{0,\cdots,2^{N}-1\}$. Since the encoding vectors are equally likely, the aforementioned decoding scheme is equal to the maximum likelihood decoding, i.e., the receiver chooses the number from $\mathcal{B}$ such that the probability $\Pr\{\textbf{z}_{i}, \textbf{r}_{i}, \textbf{r}_{i+1}| \textbf{b}_{i}\}$ is maximized. Particularly, we can express the probability density function of the channel output given the channel input as follows:
\begin{align}\label{eq:pdf_y}
	f_{\textbf{y}_{i}|\textbf{x}_{i}}(\textbf{y}_{i}|\textbf{x}_{i}) = \frac{1}{\pi^{2L+T}\det\{\Sigma_{i}\}}\exp\left\{-\textbf{y}_{i}^{\dagger}\Sigma^{-1}_{i}\textbf{y}_{i}\right\}, 
\end{align}
where $\textbf{y}_{i} = [\textbf{r}^{\dagger}_{i+1}\;\textbf{z}^{\dagger}_{i}\;\textbf{r}^{\dagger}_{i}]^{\dagger}$, $\textbf{x}_{i} = [\textbf{p}^{\dagger}_{i+1}\; \textbf{s}^{\dagger}_{i}\; \textbf{p}^{\dagger}_{i}]^{\dagger}$, and $\Sigma_{i}$ is the covariance matrix, which is expressed as $\Sigma_{i} = \diag\{\textbf{x}_{i}\}\Sigma_{\textbf{h}_{i}}\diag\{\textbf{x}_{i}\}^{\dagger} + \sigma_{w}^{2}\textbf{I}_{(2L+T)\times(2L+T)}$. Here, $\diag\{\cdot\}$ is the diagonal matrix, and $\Sigma_{\textbf{h}_{i}}$ is the channel covariance matrix, i.e., $\Sigma_{\textbf{h}_{i}} = \mathbb{E}\left\{\textbf{h}_{i}\textbf{h}_{i}^{\dagger}\right\}$, where $\textbf{h}_{i} = [h^{\dagger}((i+1)(L+T)+L-1),\cdots,h^{\dagger}(i(L+T))]^{\dagger}$ is the vector of fading coefficients corresponding to the transmitted vector, $\textbf{x}_{i}$. Moreover, $\textbf{I}$ is the identity matrix with the given size. We also remark that given an input, $\textbf{x}_{i}$, the expected channel output is zero-mean since the fading gains are zero-mean and circularly-symmetric complex. We further recall that $\textbf{s}_{i}$ is the modulator output given that the encoder input is $\textbf{b}_{i}$ as seen in Fig. \ref{fig:fig_1}. Now, calculating the log-likelihood function associated with (\ref{eq:pdf_y}) and removing the constants, one can show that the ideal receiver output becomes $\widehat{\textbf{b}}_{i} = \arg\min\left\{\textbf{b}_{i}:\textbf{y}_{i}^{\dagger}\Sigma^{-1}_{i}\textbf{y}_{i}\right\}$.

Recall that the pilot symbols are known by the receiver. However, under most conditions, it is not straightforward to find the maximum of the likelihood function, and numerical techniques become necessary. Moreover, the complexity of finding the maximum generally increases with increasing $N$. Here, we consider the maximum likelihood decoder as a basis to understand the performance limits that our proposed methods can reach. Moreover, along with the maximum likelihood decoding, we consider two different models. In the first model, the receiver initially estimates the channel fading gain, $h(t)$, and then successively performs demodulation and decoding. In the second model, we propose a machine learning tool, namely \ac{SVM}, that replaces the channel estimation, demodulation and decoding processes in order to perform all of them jointly.

\begin{figure}
	\begin{center}
		\begin{tikzpicture}[
    >=stealth',
    black!50,
    line width=0.7mm*\scalePicture,
    text=black,
    font = \small,
    every new ->/.style = {shorten >=1pt*\scalePicture},
    graphs/every graph/.style = {edges=rounded corners},
    block/.style ={font = \small, rectangle, draw=blue!60!black, thick, fill = white, text width=3em*\scalePicture, align=center, rounded corners, minimum height=3.6em*\scalePicture},
    point/.style={rectangle, minimum height=1.0em*\scalePicture, text width=3em*\scalePicture, draw = white, fill = white},
    cross/.style={cross out, draw=black, minimum size=2.0em*\scalePicture), inner sep=3pt*\scalePicture, outer sep=2pt*\scalePicture},
    cross/.default={1pt*\scalePicture}
    ]
    
    \coordinate (start) at (0pt*\scalePicture,0pt*\scalePicture) {}; 
    \node [rotate = 0, text width=4em*\scalePicture, align=center] (nodetext1) at (start) {$\tpl{\mathbf{\widehat{b}}_{i}}$};
    
    \coordinate (c1) at ($(start) + (180:45pt*\scalePicture)$) {};
    \draw [black, solid, line width=0.6mm*\scalePicture] (c1) [->] -- ($(start) + (180:10pt*\scalePicture)$);
    
    \coordinate (c2) at ($(c1)!0.30!(start) + (90:18pt*\scalePicture)$) {};
    \node [rotate = 0, text width=2em*\scalePicture, align=center] (nodetext2) at (c2) {\color{orange}{$N$ bits}};
    
    \coordinate (c3) at ($(c1) + (180:25pt*\scalePicture)$) {};
    \node [block, draw = black!20!Cyan, fill = white!60!Cyan, text width=4em*\scalePicture, minimum height=1.2em*\scalePicture, rotate=0, align = center] (Decoder) at (c3) {Decoder};
    
    \coordinate (c4) at ($(c3) + (180:60pt*\scalePicture)$) {};
    \draw [black, solid, line width=0.6mm*\scalePicture] (c4) [->] -- ($(c3) + (180:25pt*\scalePicture)$);
    
    \coordinate (c5) at ($(c4)!0.25!(c3) + (90:18pt*\scalePicture)$) {};
    \node [rotate = 0, text width=2em*\scalePicture, align=center] (nodetext2) at (c5) {\color{orange}{$M$ bits}};
    
    \coordinate (c6) at ($(c4) + (180:38pt*\scalePicture)$) {};
    \node [block, draw = black!20!Cyan, fill = white!60!Cyan, text width=7em*\scalePicture, minimum height=1.2em*\scalePicture, rotate=0] (Demodulator) at (c6) {Demodulator};
    
    \coordinate (c7) at ($(c6) + (180:100pt*\scalePicture)$) {};
    \draw [black, solid, line width=0.6mm*\scalePicture] (c7) [->] -- ($(c6) + (180:40pt*\scalePicture)$);
    
    \coordinate (c8) at ($(c7) + (270:80pt*\scalePicture)$) {};
    \coordinate (c9) at ($(c7)!0.5!(c8)$) {};
    \coordinate (c10) at ($(c9) + (180:38pt*\scalePicture)$) {};
    \coordinate (c11) at ($(c6) + (270:80pt*\scalePicture)$) {};
    \coordinate (c12) at ($(c3) + (270:80pt*\scalePicture)$) {};
    \coordinate (c13) at ($(c1)!0.5!(start)$) {};
    \coordinate (c14) at ($(c13) + (270:80pt*\scalePicture)$) {};
    \coordinate (c15) at ($(c13)!0.5!(c14)$) {};
    
    \draw [black, solid, line width=0.6mm*\scalePicture] (c8) [->] -- ($(c11) + (180:35pt*\scalePicture)$);
    
    \node [block, draw = black!20!Cyan, fill = white!60!Cyan, text width=9em*\scalePicture, minimum height=2.5em*\scalePicture, rotate=270] (node1) at ($(c9)$) {De-multiplexer};
    
    \draw [black, solid, line width=0.6mm*\scalePicture] (c10) [->] -- ($(c9) + (180:13pt*\scalePicture)$);
    
    \draw [black, solid, line width=0.6mm*\scalePicture] (c11) [->] -- ($(c6) + (270:10pt*\scalePicture)$);
    
    \draw [Plum, dash pattern={on 1pt off 1pt}, line width=0.6mm*\scalePicture] (c12) [->] -- ($(c11) + (0:35pt*\scalePicture)$);
    
    \draw [Plum, dash pattern={on 1pt off 1pt}, line width=0.6mm*\scalePicture] (c13) -- (c14) [->] -- ($(c12) + (0:30pt*\scalePicture)$);
    
    \draw [black, dash pattern={on 0.5pt off 0.5pt}, line width=0.5mm*\scalePicture] ($(c15) + (180:20pt*\scalePicture)$) -- ($(c15) + (0:20pt*\scalePicture)$);
    
    \node [block, draw = black!20!Cyan, fill = white!60!Cyan, text width=6em*\scalePicture, minimum height=1.2em*\scalePicture, rotate=0] (ChannelEstimation) at (c11) {Channel Estimation};
    
    \node [block, draw = Plum, dash pattern={on 1pt off 1pt}, fill = white!60!Cyan, text width=5em*\scalePicture, minimum height=1.2em*\scalePicture, rotate=0] (Feedback) at (c12) {Feedback};
    
    \coordinate (RPS) at ($(c8)!0.4!(c11) + (270:20pt*\scalePicture)$) {};
    \node [rotate = 0, text width=12em*\scalePicture, align=center] (nodetextRPS) at (RPS) {\color{orange}{Received\\Pilot Symbols}};
    
    \coordinate (ri) at ($(c8)!0.4!(c11) + (90:12pt*\scalePicture)$) {};
    \node [rotate = 0, text width=4em*\scalePicture, align=center] (nodetextri) at (ri) {$\tpl{\mathbf{r}_{i+1}}, \tpl{\mathbf{r}_{i}}$};
    
    \coordinate (RDS) at ($(c7)!0.4!(c6) + (90:20pt*\scalePicture)$) {};
    \node [rotate = 0, text width=12em*\scalePicture, align=center] (nodetextRDS) at (RDS) {\color{orange}{Received Data\\Symbols}};
    
    \coordinate (zi) at ($(c7)!0.3!(c6) + (270:12pt*\scalePicture)$) {};
    \node [rotate = 0, text width=4em*\scalePicture, align=center] (nodetextzi) at (zi) {$\tpl{\mathbf{z}_{i}}$};
    
    \coordinate (fge) at ($(c11)!0.53!(c6) + (0:26pt*\scalePicture)$) {};
    \node [rotate = 0, text width=6em*\scalePicture, align=center] (nodetextfge) at (fge) {\color{orange}{Fading Gain Estimate}};
    
    \coordinate (yt) at ($(c10)!0.3!(c9) + (90:11pt*\scalePicture)$) {};
    \node [rotate = 0, text width=4em*\scalePicture, align=center] (nodetextyt) at (yt) {$\color{orange}{y(t)}$};
    
    \coordinate (bi1) at ($(c14) + (90:11pt*\scalePicture) + (0:17pt*\scalePicture)$) {}; 
    \node [rotate = 0, text width=4em*\scalePicture, align=center] (nodetext1) at (bi1) {$\tpl{\mathbf{\widehat{b}}_{i-1}}$};
    
    \coordinate (si1) at ($(c11)!0.6!(c12) + (270:15pt*\scalePicture)$) {}; 
    \node [rotate = 0, text width=4em*\scalePicture, align=center] (nodetext1) at (si1) {$\tpl{\mathbf{\widehat{s}}_{i-1}}$};
    
    \coordinate (c16) at ($(c8) + (0:30pt*\scalePicture) + (270:70pt*\scalePicture)$) {};
    \coordinate (c17) at ($(c16) + (0:40pt*\scalePicture)$) {};
    \draw [black, solid, line width=0.6mm*\scalePicture] (c16) -- (c17);
    \node [rotate = 0, text width=8em*\scalePicture, align=center] (nodetextinfo1) at ($(c17) + (0:120pt*\scalePicture)$) {: Technique 1};
    
    \coordinate (c18) at ($(c16) + (270:20pt*\scalePicture)$) {};
    \coordinate (c19) at ($(c17) + (270:20pt*\scalePicture)$) {};
    \draw [black, solid, line width=0.6mm*\scalePicture] (c18) -- (c19);
    \node [rotate = 0, text width=1em*\scalePicture, align=center] (nodetextinfoplut) at ($(c19) + (0:10pt*\scalePicture)$) {+};
    \draw [Plum, dash pattern={on 1pt off 1pt}, line width=0.6mm*\scalePicture] ($(c19) + (0:20pt*\scalePicture)$) -- ($(c19) + (0:60pt*\scalePicture)$);
    \node [rotate = 0, text width=8em*\scalePicture, align=center] (nodetextinfo2) at ($(c19) + (0:120pt*\scalePicture)$) {: Technique 2};    
\end{tikzpicture}
		\caption{Conventional receiver.}\label{fig:fig_4}
	\end{center}
	\vspace{-0.5cm}
\end{figure}
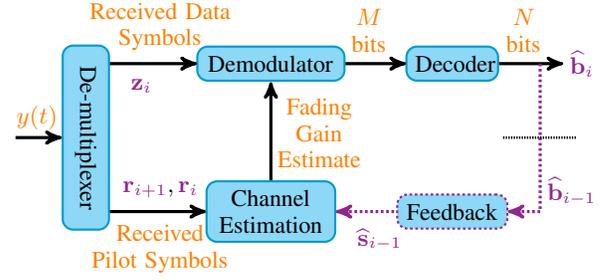
\paragraph{Model 1} We consider the receiver model given in Fig. \ref{fig:fig_4}. In this model, we consider two different channel estimation techniques, namely the minimum mean-square error (MMSE) and iterative Kalman filter estimators. As seen in Fig. \ref{fig:fig_4}, both the techniques use pilot symbols for channel estimation.

\emph{Technique 1 (MMSE)}: Given $\textbf{r}_{i}$ and $\textbf{r}_{i+1}$, we can express the channel gain estimate as $\widehat{\textbf{h}}_{i} = \mathbb{E}\left\{\textbf{h}_{i}|\textbf{r}_{i},\textbf{r}_{i+1}\right\}$, where $\widehat{\textbf{h}}_{i} = [\widehat{h}^{\dagger}((i+1)(L+T)+L-1), \cdots, \widehat{h}^{\dagger}(i(L+T))]^{\dagger}$ is the vector of channel fading estimates. Recall that $\textbf{h}_{i}$ is defined in Sec. \ref{para:idealReceiver}. Notice that the first and last $L$ elements of $\textbf{h}_{i}$ and $\widehat{\textbf{h}}_{i}$ correspond to the pilot symbols transmitted in the $(i+1)^{\text{th}}$ and $i^{\text{th}}$ time frames, and the $T$ elements in between them correspond to the data symbols transmitted in the $i^{\text{th}}$ frame. When $\textbf{h}_{i}$ is jointly Gaussian with $\textbf{r}_{i+1}$ and $\textbf{r}_{i}$, i.e., their joint distribution is a multivariate Gaussian distribution, the MMSE estimator becomes the linear MMSE (L-MMSE) estimator. Then, we can easily express the channel estimate, $\widehat{\textbf{h}}_{i}$, as follows:
\begin{equation}\label{eq:linear_channel_estimate}
\widehat{\textbf{h}}_{i} = \mathbb{E}\left\{\textbf{h}_{i}\right\} + \Sigma_{\textbf{h}_{i}\textbf{\v{r}}_{i}}\Sigma_{\textbf{\v{r}}_{i}}^{-1}\left(\textbf{\v{r}}_{i}-\mathbb{E}\left\{\textbf{\v{r}}_{i}\right\}\right),
\end{equation}
where $\Sigma_{\textbf{h}_{i}\textbf{\v{r}}_{i}} = \mathbb{E}\left\{(\textbf{h}_{i} - \mathbb{E}\left\{\textbf{h}_{i}\right\})\left(\textbf{\v{r}}_{i} - \mathbb{E}\left\{\textbf{\v{r}}_{i}\right\}\right)^{\dagger}\right\}$ and $\Sigma_{\textbf{\v{r}}_{i}} = \mathbb{E}\left\{\left(\textbf{\v{r}}_{i} - \mathbb{E}\left\{\textbf{\v{r}}_{i}\right\}\right)\left(\textbf{\v{r}}_{i} - \mathbb{E}\left\{\textbf{\v{r}}_{i}\right\}\right)^{\dagger}\right\}$ are the cross-covariance matrix between $\textbf{h}_{i}$ and $\textbf{\v{r}}_{i}$ and the auto-covariance matrix of $\textbf{\v{r}}_{i}$, respectively, and $\textbf{\v{r}}_{i} = [\textbf{r}_{i+1}^{\dagger}\;\textbf{r}_{i}^{\dagger}]^{\dagger}$. Here, we also note that because the channel is flat fading, the estimation in (\ref{eq:linear_channel_estimate}) is equivalent to the channel estimation obtained by implementing the semi-bling L-MMSE estimator \cite{dong2003optimal}.  

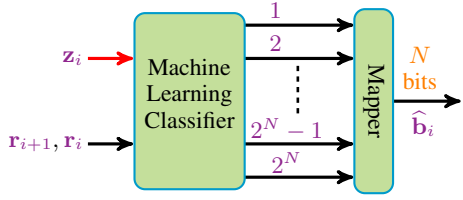
\begin{figure}
	\begin{center}
		\begin{tikzpicture}[
    >=stealth',
    black!50,
    line width=0.7mm*\scalePicture,
    text=black,
    font = \small,
    every new ->/.style = {shorten >=1pt*\scalePicture},
    graphs/every graph/.style = {edges=rounded corners},
    block/.style ={font = \small, rectangle, draw=blue!60!black, thick, fill = white, text width=3em*\scalePicture, align=center, rounded corners, minimum height=3.6em*\scalePicture},
    point/.style={rectangle, minimum height=1.0em*\scalePicture, text width=3em*\scalePicture, draw = white, fill = white},
    cross/.style={cross out, draw=black, minimum size=2.0em*\scalePicture), inner sep=3pt*\scalePicture, outer sep=2pt*\scalePicture},
    cross/.default={1pt*\scalePicture}
    ]
    \matrix[column sep=7mm*\scalePicture, row sep=5mm*\scalePicture]
    {
        \node [point] (p11) {}; & \node [point] (p12) {}; & \node [point] (p13) {}; & \node [point] (p14) {};\\
        \node [point] (p21) {}; & \node [point] (p22) {}; & \node [point] (p23) {}; & \node [point] (p24) {};\\
        \node [point] (p31) {}; & \node [point] (p32) {}; & \node [point] (p33) {}; & \node [point] (p34) {};\\
        \node [point] (p41) {}; & \node [point] (p42) {}; & \node [point] (p43) {}; & \node [point] (p44) {};\\
    };
    

    \coordinate (t12) at ($(p21)!0.5!(p31) + (180:13.5pt*\scalePicture)$) {};
    \coordinate (t21) at ($(p21) + (0:10pt*\scalePicture) + (90:10pt*\scalePicture)$) {};
    \coordinate (t22) at ($(p22) + (180:27pt*\scalePicture) + (90:10pt*\scalePicture)$) {};
    \draw [red, solid, line width=0.6mm*\scalePicture] (t21) [->] -- (t22);
    
    \coordinate (t31) at ($(p31) + (0:10pt*\scalePicture) + (270:10pt*\scalePicture)$) {};
    \coordinate (t32) at ($(p32) + (180:27pt*\scalePicture) + (270:10pt*\scalePicture)$) {};
    \draw [black, solid, line width=0.6mm*\scalePicture] (t31) [->] -- (t32);
    
    \coordinate (t41) at ($(p22)!0.5!(p32) + (0:110pt*\scalePicture)$) {};
    \coordinate (t42) at ($(p22)!0.5!(p32) + (0:150pt*\scalePicture)$) {};
    \draw [black, solid, line width=0.6mm*\scalePicture] (t41) [->] -- (t42);
    
    
    \coordinate (text2) at ($(t21) + (180:8pt*\scalePicture)$) {};
    \node [rotate = 0, text width=4em*\scalePicture, align=center] (nodetext2) at (text2) {$\tpl{\mathbf{z}_{i}}$};
    
    \coordinate (text3) at ($(t31) + (180:23pt*\scalePicture)$) {};
    \node [rotate = 0, text width=4em*\scalePicture, align=center] (nodetext3) at (text3) {$\tpl{\mathbf{r}_{i+1}}, \tpl{\mathbf{r}_{i}}$};
        
    \coordinate (text4) at ($(t41)!0.45!(t42) + (90:18pt*\scalePicture)$) {};
    \node [rotate = 0, text width=2em*\scalePicture, align=center] (nodetext4) at (text4) {\color{orange}{$N$ bits}};
    
    \coordinate (text5) at ($(t41)!0.5!(t42) + (270:12pt*\scalePicture)$) {};
    \node [rotate = 0, text width=4em*\scalePicture, align=center] (nodetext5) at (text5) {$\tpl{\mathbf{\widehat{b}}_{i}}$};
    
    \coordinate (a1) at ($(t22) + (0:59pt*\scalePicture) + (90:18pt*\scalePicture)$) {};
    \coordinate (a11) at ($(a1) + (0:60pt*\scalePicture)$) {};
    \draw [black, solid, line width=0.6mm*\scalePicture] (a1) [->] -- (a11);
    \node [rotate = 0, text width=4em*\scalePicture, align=center] (nodetext6) at ($(a1)!0.3!(a11) + (90:8pt*\scalePicture)$) {$\tpl{1}$};
    
    \coordinate (a2) at ($(a1) + (270:18pt*\scalePicture)$) {};
    \coordinate (a21) at ($(a11) + (270:18pt*\scalePicture)$) {};
    \draw [black, solid, line width=0.6mm*\scalePicture] (a2) [->] -- (a21);
    \node [rotate = 0, text width=4em*\scalePicture, align=center] (nodetext7) at ($(a2)!0.3!(a21) + (90:8pt*\scalePicture)$) {$\tpl{2}$};


    \coordinate (a3) at ($(t32) + (0:59pt*\scalePicture) + (270:18pt*\scalePicture)$) {};
    \coordinate (a31) at ($(a3) + (0:60pt*\scalePicture)$) {};
    \draw [black, solid, line width=0.6mm*\scalePicture] (a3) [->] -- (a31);
    \node [rotate = 0, text width=4em*\scalePicture, align=center] (nodetext7) at ($(a3)!0.4!(a31) + (90:8pt*\scalePicture)$) {$\tpl{2^{N}}$};
    
    \coordinate (a4) at ($(a3) + (90:18pt*\scalePicture)$) {};
    \coordinate (a41) at ($(a31) + (90:18pt*\scalePicture)$) {};
    \draw [black, solid, line width=0.6mm*\scalePicture] (a4) [->] -- (a41);
    \node [rotate = 0, text width=4em*\scalePicture, align=center] (nodetext7) at ($(a4)!0.4!(a41) + (90:8pt*\scalePicture)$) {$\tpl{2^{N}-1}$};
    
    
    \coordinate (dot1) at ($(a2)!0.5!(a21) + (270:4pt*\scalePicture)$) {};
    \coordinate (dot2) at ($(a4)!0.5!(a41) + (90:14pt*\scalePicture)$) {};
    \draw [black, dash pattern={on 2pt off 2pt}, line width=0.5mm*\scalePicture] (dot1) -- (dot2);
    
    
    \coordinate (Receiver) at ($(p22)!0.5!(p32) + (0:3.5pt*\scalePicture)$) {};
    \node [block, draw = black!20!Cyan, fill = LimeGreen!50!White, text width=5em*\scalePicture, minimum height=2.5em*\scalePicture, rotate=0] (node2) at ($(Receiver)$) {\text{ }\\\text{ }\\Machine Learning Classifier\\\text{ }\\\text{ }};
    
    \coordinate (mapper) at ($(a21)!0.5!(a41) + (0:11pt*\scalePicture)$) {};
    \node [block, draw = black!20!Cyan, fill = LimeGreen!50!White, text width=9em*\scalePicture, minimum height=2.0em*\scalePicture, rotate=270] (nodeMapper) at (mapper) {Mapper};
\end{tikzpicture}
		\caption{Model 2, Technique 1.}\label{fig:fig_5}
	\end{center}
	\vspace{-0.5cm}
\end{figure}
\emph{Technique 2 (Kalman)}: The receiver performs the channel estimation in two steps, i.e., the estimation process is composed of the prediction and update of the channel fading gains. Particularly, using the defined fading process, we define the state transition as follows:
\begin{align*}
\textbf{h}_{i} = &\begin{bmatrix}\boldsymbol{\alpha}_{v}& \textbf{0}_{(L+T)\times(2L+T-1)}\\\textbf{I}_{L\times L}&\textbf{0}_{L\times(L+T)}\end{bmatrix}\textbf{h}_{i-1}+\begin{bmatrix}\boldsymbol{\alpha}_{m}\\\textbf{0}_{L\times(L+T)}\end{bmatrix}\boldsymbol{\beta}_{i},
\end{align*}
where $\boldsymbol{\alpha} = [\alpha^{L+T},\cdots,\alpha^{2},\alpha]^{\dagger}$, $\boldsymbol{\beta}_{i} = [\beta^{\dagger}((i+1)(L+T)+L-1)\; \cdots \; \beta^{\dagger}(i(L+T)+L)]^{\dagger}$, $\textbf{0}$ is the zero matrix with the defined size, and
\begin{align*}
	\boldsymbol{\alpha}_{m} = \begin{bmatrix}1&\alpha&\cdots&\alpha^{L+T-1}\\0&1&\cdots&\alpha^{L+T-2}\\\vdots&\vdots&\vdots&\vdots\\0&0&\cdots&1\end{bmatrix}.
\end{align*}
Note that $\beta(t)$ is the state noise given in (\ref{eq:channel_model}). Furthermore, we define the observation process as follows:
\begin{align}
\textbf{\v{r}}_{i} = & \begin{bmatrix}\diag\{\textbf{p}_{i+1}\}&\textbf{0}_{L\times(L+T)}\\\textbf{0}_{L\times(L+T)}&\diag\{\textbf{p}_{i}\}\end{bmatrix}\textbf{h}_{i}\nonumber \\&\hspace{2.3cm}+ \begin{bmatrix}w((i+1)(L+T)+L-1)\\\vdots\\w((i+1)(L+T))\\w(i(L+T)+L-1)\\\vdots\\w(i(L+T))\end{bmatrix},\label{eq:observation}
\end{align}
where $w(t)$ is the additive thermal noise given in (\ref{eq:channel_input_output}), and $\diag\{\textbf{p}_{i+1}\}$ and $\diag\{\textbf{p}_{i}\}$ are $L\times L$ diagonal matrices. Having the state transition and observation processes, one can easily compute the Kalman gain matrix, and predict and update the channel fading estimate accordingly. Due to the space limit, we refer interested readers to \cite{bishop2001introduction} for more details on the Kalman filter estimator.

\emph{Demodulation and Decoding}: Now, following any of the aforementioned channel estimation process, i.e., \emph{MMSE} or \emph{iterative Kalman} estimator, we can express the channel input-output relation during the data transmission as follows:
\begin{equation}\label{eq:channel_input_output_2}
z_{i}(t) = \widehat{h}(t)s_{i}(t) + \widetilde{h}(t)s_{i}(t) + w(t).
\end{equation}
Treating the channel estimate error as another source of additive noise, the receiver demodulates the received symbols:
\begin{equation}\label{eq:channel_input_output_3}
\frac{\widehat{h}^{\dagger}(t)z_{i}(t)}{|\widehat{h}(t)|^{2}} = s_{i}(t) + \frac{\widehat{h}^{\dagger}(t)\widetilde{h}(t)x_{t} + \widehat{h}^{\dagger}(t)w(t)}{|\widehat{h}(t)|^{2}}.
\end{equation}
Subsequently, using the demodulated data, the receiver decodes the transmitted data, $\widehat{\textbf{b}}_{i}$.

\paragraph{Model 2} The receiver performs the channel estimation, demodulation and decoding processes jointly using an \ac{SVM}-based classification tool. Different from \emph{Model 1}, we initially train the receiver in \emph{Model 2} because \acp{SVM} are supervised learning algorithms. Thus, we need a packet of training data. One can obtain this data either by implementing the scenario in real-time and save the transmitted bits and the channel outputs, or by simulating the channel based on known models because we know that available channel models can reflect the real channels very closely. We define the input (feature vector\footnote{In the literature, the term \emph{feature vector} is mostly used to define the input to an \ac{SVM}-based classifier. Throughout the paper, we use \emph{input} and \emph{feature vector} interchangeably.}) to the \ac{SVM}-based receiver as $[\textbf{r}_{i+1}\;\textbf{r}_{i}\;\textbf{z}_{i}]$ and propose two different techniques. The \ac{SVM}-based decoder projects the feature vector to a higher dimensional space and find hyper-planes that separate given encoding vectors. Specifically, the \ac{SVM}-based receiver obtains support vectors during the training phase implementing certain \emph{kernel} functions, and use them in order to classify the input. In order to have a better intuition, one can think of each support vector as a dimension and the kernel function as the projector of the input vector onto the support vectors. In this paper, we run non-linear classifiers; therefore, we consider two kernel functions for the SVM-based receiver, namely the polynomial function with degree $2$ and the Gaussian radial basis function with auto-scale. We finally note that although we use an SVM-based receiver in our analysis, one can implement other supervised learning algorithms.
\begin{figure}
	\begin{center}
		\begin{tikzpicture}[
    >=stealth',
    black!50,
    line width=0.7mm*\scalePicture,
    text=black,
    font = \small,
    every new ->/.style = {shorten >=1pt*\scalePicture},
    graphs/every graph/.style = {edges=rounded corners},
    block/.style ={font = \small, rectangle, draw=blue!60!black, thick, fill = white, text width=3em*\scalePicture, align=center, rounded corners, minimum height=3.6em*\scalePicture},
    point/.style={rectangle, minimum height=1.0em*\scalePicture, text width=3em*\scalePicture, draw = white, fill = white},
    cross/.style={cross out, draw=black, minimum size=2.0em*\scalePicture), inner sep=3pt*\scalePicture, outer sep=2pt*\scalePicture},
    cross/.default={1pt*\scalePicture}
    ]
    
    \coordinate (ML1) at (0pt*\scalePicture,0pt*\scalePicture) {};
    \coordinate (ML2) at ($(ML1) + (270:55pt*\scalePicture)$) {};
    \coordinate (MLdots) at ($(ML2) + (270:55pt*\scalePicture)$) {};
    \coordinate (MLN) at ($(MLdots) + (270:55pt*\scalePicture)$) {};
    
    
    \coordinate (z1) at ($(ML1) + (180:120pt*\scalePicture) + (90:13pt*\scalePicture)$) {};
    \coordinate (z2) at ($(z1) + (0:79pt*\scalePicture)$) {};
    \draw [red, solid, line width=0.6mm*\scalePicture] (z1) [->] -- (z2);
    
    \coordinate (z3) at ($(z1) + (0:29pt*\scalePicture)$) {};
    \coordinate (z4) at ($(z3) + (270:55pt*\scalePicture)$) {};
    \coordinate (z5) at ($(z2) + (270:55pt*\scalePicture)$) {};
    \draw [red, solid, line width=0.6mm*\scalePicture] (z3) -- (z4) [->] -- (z5);
    
    \coordinate (z6) at ($(z4) + (270:110pt*\scalePicture)$) {};
    \coordinate (z7) at ($(z5) + (270:110pt*\scalePicture)$) {};
    \draw [red, solid, line width=0.6mm*\scalePicture] (z4) -- (z6) [->] -- (z7);
    
    \coordinate (r1) at ($(ML1) + (180:120pt*\scalePicture) + (270:13pt*\scalePicture)$) {};
    \coordinate (r2) at ($(r1) + (0:79pt*\scalePicture)$) {};
    \draw [black, solid, line width=0.6mm*\scalePicture] (r1) [->] -- (r2);
    \coordinate (r3) at ($(r1) + (0:19pt*\scalePicture)$) {};
    \coordinate (r4) at ($(r3) + (270:55pt*\scalePicture)$) {};
    \coordinate (r5) at ($(r2) + (270:55pt*\scalePicture)$) {};
    \draw [black, solid, line width=0.6mm*\scalePicture] (r3) -- (r4) [->] -- (r5);
    \coordinate (r6) at ($(r4) + (270:110pt*\scalePicture)$) {};
    \coordinate (r7) at ($(r5) + (270:110pt*\scalePicture)$) {};
    \draw [black, solid, line width=0.6mm*\scalePicture] (r4) -- (r6) [->] -- (r7);

    \coordinate (mout1) at ($(ML1) + (0:40pt*\scalePicture)$) {};
    \coordinate (mout12) at ($(ML1) + (0:80pt*\scalePicture)$) {};
    \draw [black, solid, line width=0.6mm*\scalePicture] (mout1) [->] -- (mout12);
    
    \coordinate (mout2) at ($(ML2) + (0:40pt*\scalePicture)$) {};
    \coordinate (mout22) at ($(ML2) + (0:80pt*\scalePicture)$) {};
    \draw [black, solid, line width=0.6mm*\scalePicture] (mout2) [->] -- (mout22);
    
    \coordinate (moutdots) at ($(MLdots) + (0:60pt*\scalePicture)$) {};
    \draw [black, dash pattern={on 2pt off 2pt}, line width=0.5mm*\scalePicture] ($(moutdots) + (90:37pt*\scalePicture)$) -- ($(moutdots) + (270:23pt*\scalePicture)$);
    
    \coordinate (moutN) at ($(MLN) + (0:40pt*\scalePicture)$) {};
    \coordinate (moutN2) at ($(MLN) + (0:80pt*\scalePicture)$) {};
    \draw [black, solid, line width=0.6mm*\scalePicture] (moutN) [->] -- (moutN2);
    
    
    \coordinate (zText) at ($(z1) + (180:7pt*\scalePicture)$) {};
    \node [rotate = 0, text width=4em*\scalePicture, align=center] (nodetextZ) at (zText) {$\tpl{\mathbf{z}_{i}}$};
    
    \coordinate (rText) at ($(r1) + (180:22pt*\scalePicture)$) {};
    \node [rotate = 0, text width=4em*\scalePicture, align=center] (nodetextR) at (rText) {$\tpl{\mathbf{r}_{i+1}}, \tpl{\mathbf{r}_{i}}$};
    
    \coordinate (text1) at ($(mout1)!0.5!(mout12) + (90:10pt*\scalePicture)$) {};
    \node [rotate = 0, text width=4em*\scalePicture, align=center] (nodetext1) at (text1) {\tpl{1}};
    
    \coordinate (text2) at ($(mout2)!0.5!(mout22) + (90:10pt*\scalePicture)$) {};
    \node [rotate = 0, text width=4em*\scalePicture, align=center] (nodetext2) at (text2) {\tpl{2}};
    
    \coordinate (textN) at ($(moutN)!0.5!(moutN2) + (90:10pt*\scalePicture)$) {};
    \node [rotate = 0, text width=4em*\scalePicture, align=center] (nodetext1) at (textN) {$\tpl{N}$};
    
    \coordinate (mapper) at ($(mout12)!0.5!(moutN2) + (0:11pt*\scalePicture)$) {};
    \node [block, draw = black!20!Cyan, fill = LimeGreen!50!White, text width=20em*\scalePicture, minimum height=2.0em*\scalePicture, rotate=270] (nodeMapper) at (mapper) {Mapper};
    
    \coordinate (bi1) at ($(mapper) + (0:11pt*\scalePicture)$) {};
    \coordinate (bi2) at ($(mapper) + (0:41pt*\scalePicture)$) {};
    \draw [black, solid, line width=0.6mm*\scalePicture] (bi1) [->] -- (bi2);
    
    \coordinate (textbi) at ($(bi1)!0.5!(bi2) + (90:12pt*\scalePicture)$) {};
    \node [rotate = 0, text width=4em*\scalePicture, align=center] (nodetextbi) at (textbi) {$\tpl{\mathbf{\widehat{b}}_{i}}$};
    
    \node [block, draw = black!20!Cyan, fill = LimeGreen!50!White, text width=7em*\scalePicture, minimum height=2.5em*\scalePicture, rotate=0] (nodeML1) at (ML1) {Machine Learning Classifier 1};
    \node [block, draw = black!20!Cyan, fill = LimeGreen!50!White, text width=7em*\scalePicture, minimum height=2.5em*\scalePicture, rotate=0] (nodeML2) at (ML2) {Machine Learning Classifier 2};
    \draw [black, dash pattern={on 2pt off 2pt}, line width=0.5mm*\scalePicture] ($(MLdots) + (90:20pt*\scalePicture)$) -- ($(MLdots) + (270:20pt*\scalePicture)$);
    \node [block, draw = black!20!Cyan, fill = LimeGreen!50!White, text width=7em*\scalePicture, minimum height=2.5em*\scalePicture, rotate=0] (nodeMLN) at (MLN) {Machine Learning Classifier $N$};
\end{tikzpicture}
		\caption{Model 2, Technique 2.}\label{fig:fig_6}
	\end{center}
	\vspace{-0.5cm}
\end{figure}
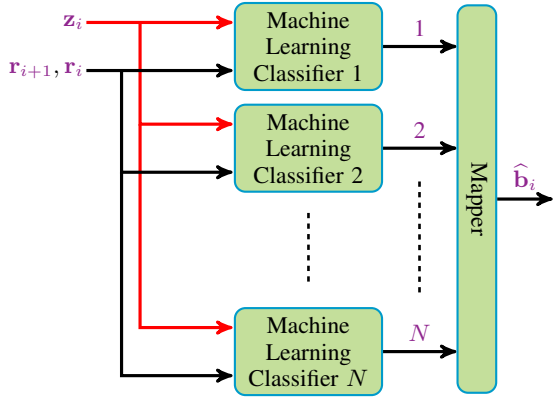

\begin{figure}[t]
	\begin{center}
%
%
\definecolor{mycolor1}{rgb}{0.85000,0.32500,0.09800}%
\definecolor{mycolor2}{rgb}{0.92900,0.69400,0.12500}%
\definecolor{mycolor3}{rgb}{0.49400,0.18400,0.55600}%
\definecolor{mycolor4}{rgb}{0.46600,0.67400,0.18800}%
\definecolor{mycolor5}{rgb}{0.63500,0.07800,0.18400}%
\begin{tikzpicture}

\begin{axis}[%
font = \footnotesize,
width=4.602in*\scaleFigure,
height=3.552in*\scaleFigure,
at={(0.772in*\scaleFigure,0.495in*\scaleFigure)},
scale only axis,
xmin=5,
xmax=30,
xlabel style={font=\color{white!15!black}\footnotesize},
xlabel={Signal-to-noise ratio, SNR (dB)},
ymin=0.03,
ymax=0.25,
ylabel style={font=\color{white!15!black}\footnotesize},
ylabel={Bit-error probability (BER)},
axis background/.style={fill=white},
title style={font={\footnotesize}},
title={$\alpha =0.95$},
legend style={at={(0.5735*\scaleFigure,0.9748*\scaleFigure)}, anchor=south west, legend cell align=left, align=left, draw=white!15!black}
]
\addplot [color=mycolor1, dashdotted, line width=3.0pt, mark=asterisk, mark options={solid, mycolor1}]
  table[row sep=crcr]{%
5	0.23681\\
10	0.1719775\\
15	0.1413375\\
20	0.129615\\
25	0.12751\\
30	0.1263675\\
};
\addlegendentry{\scriptsize{Model 1, Tech 1}}

\addplot [color=mycolor2, line width=2.0pt, mark=square, mark options={solid, mycolor2}]
  table[row sep=crcr]{%
5	0.2345095\\
10	0.172621\\
15	0.1420415\\
20	0.13045325\\
25	0.126567\\
30	0.12455225\\
};
\addlegendentry{\scriptsize{Model 1, Tech 2}}

\addplot [color=mycolor3, line width=2.0pt, mark=x, mark options={solid, mycolor3}]
  table[row sep=crcr]{%
5	0.220155\\
10	0.1414975\\
15	0.10214\\
20	0.08736\\
25	0.0825075\\
30	0.08044\\
};
\addlegendentry{\scriptsize{Model 2, Tech 2, Gaussian}}

\addplot [color=mycolor4, dashdotted, line width=3.0pt, mark=triangle, mark options={solid, rotate=180, mycolor4}]
  table[row sep=crcr]{%
5	0.2169275\\
10	0.1394975\\
15	0.09801\\
20	0.08166\\
25	0.075995\\
30	0.0731475\\
};
\addlegendentry{\scriptsize{Model 2, Tech 1, Gaussian}}

\addplot [color=mycolor5, line width=2.0pt, mark=o, mark options={solid, mycolor5}]
  table[row sep=crcr]{%
5	0.212645\\
10	0.137405\\
15	0.0969775\\
20	0.0800775\\
25	0.074185\\
30	0.0713425\\
};
\addlegendentry{\scriptsize{Model 2, Tech 1, Polynomial}}

\addplot [color=black, line width=1.0pt, mark=diamond, mark options={solid, black}]
  table[row sep=crcr]{%
5	0.2100935\\
10	0.13719225\\
15	0.09709725\\
20	0.07839375\\
25	0.07101625\\
30	0.0687115\\
};
\addlegendentry{\scriptsize{Maximum Likelihood Decoder}}

\end{axis}

\end{tikzpicture}%
\\
%
%
\definecolor{mycolor1}{rgb}{0.85000,0.32500,0.09800}%
\definecolor{mycolor2}{rgb}{0.92900,0.69400,0.12500}%
\definecolor{mycolor3}{rgb}{0.49400,0.18400,0.55600}%
\definecolor{mycolor4}{rgb}{0.46600,0.67400,0.18800}%
\definecolor{mycolor5}{rgb}{0.63500,0.07800,0.18400}%
\begin{tikzpicture}

\begin{axis}[%
font = \footnotesize,
width=4.602in*\scaleFigure,
height=3.552in*\scaleFigure,
at={(0.772in*\scaleFigure,0.495in*\scaleFigure)},
scale only axis,
xmin=5,
xmax=30,
xlabel style={font=\color{white!15!black}\footnotesize},
xlabel={Signal-to-noise ratio, SNR (dB)},
ymin=0.03,
ymax=0.25,
ylabel style={font=\color{white!15!black}\footnotesize},
ylabel={Bit-error probability (BER)},
axis background/.style={fill=white},
title style={font=\footnotesize},
title={$\alpha =0.97$},
legend style={at={(0.5735*\scaleFigure,0.9748*\scaleFigure)}, anchor=south west, legend cell align=left, align=left, draw=white!15!black}
]
\addplot [color=mycolor1, dashdotted, line width=3.0pt, mark=asterisk, mark options={solid, mycolor1}]
  table[row sep=crcr]{%
5	0.2185225\\
10	0.15103\\
15	0.1164875\\
20	0.102695\\
25	0.096585\\
30	0.0953375\\
};
\addlegendentry{\scriptsize{Model 1, Tech 1}}

\addplot [color=mycolor2, line width=2.0pt, mark=square, mark options={solid, mycolor2}]
  table[row sep=crcr]{%
5	0.217144\\
10	0.15134275\\
15	0.117024\\
20	0.10232725\\
25	0.096881\\
30	0.09498825\\
};
\addlegendentry{\scriptsize{Model 1, Tech 2}}

\addplot [color=mycolor3, line width=2.0pt, mark=x, mark options={solid, mycolor3}]
  table[row sep=crcr]{%
5	0.2067675\\
10	0.1300675\\
15	0.08838\\
20	0.071595\\
25	0.0648325\\
30	0.0623475\\
};
\addlegendentry{\scriptsize{Model 2, Tech 2, Gaussian}}

\addplot [color=mycolor4, dashdotted, line width=3.0pt, mark=triangle, mark options={solid, rotate=180, mycolor4}]
  table[row sep=crcr]{%
5	0.2035425\\
10	0.127815\\
15	0.08493\\
20	0.0654775\\
25	0.0573775\\
30	0.05504\\
};
\addlegendentry{\scriptsize{Model 2, Tech 1, Gaussian}}

\addplot [color=mycolor5, line width=2.0pt, mark=o, mark options={solid, mycolor5}]
  table[row sep=crcr]{%
5	0.2007\\
10	0.126975\\
15	0.08468\\
20	0.06462\\
25	0.056\\
30	0.05365\\
};
\addlegendentry{\scriptsize{Model 2, Tech 1, Polynomial}}

\addplot [color=black, line width=1.0pt, mark=diamond, mark options={solid, black}]
  table[row sep=crcr]{%
5	0.19966925\\
10	0.1262915\\
15	0.08404625\\
20	0.06288375\\
25	0.05453175\\
30	0.05148425\\
};
\addlegendentry{\scriptsize{Maximum Likelihood Decoder}}

\end{axis}
\end{tikzpicture}%
\\
%
%
\definecolor{mycolor1}{rgb}{0.85000,0.32500,0.09800}%
\definecolor{mycolor2}{rgb}{0.92900,0.69400,0.12500}%
\definecolor{mycolor3}{rgb}{0.49400,0.18400,0.55600}%
\definecolor{mycolor4}{rgb}{0.46600,0.67400,0.18800}%
\definecolor{mycolor5}{rgb}{0.63500,0.07800,0.18400}%
\begin{tikzpicture}

\begin{axis}[%
font = \footnotesize,
width=4.602in*\scaleFigure,
height=3.552in*\scaleFigure,
at={(0.772in*\scaleFigure,0.495in*\scaleFigure)},
scale only axis,
xmin=5,
xmax=30,
xlabel style={font=\color{white!15!black}\footnotesize},
xlabel={Signal-to-noise ratio, SNR (dB)},
ymin=0.03,
ymax=0.25,
ylabel style={font=\color{white!15!black}\footnotesize},
ylabel={Bit-error probability (BER)},
axis background/.style={fill=white},
title style={font=\footnotesize},
title={$\alpha =0.99$},
legend style={at={(0.5735*\scaleFigure,0.9748*\scaleFigure)}, anchor=south west, legend cell align=left, align=left, draw=white!15!black}
]
\addplot [color=mycolor1, dashdotted, line width=3.0pt, mark=asterisk, mark options={solid, mycolor1}]
  table[row sep=crcr]{%
5	0.197325\\
10	0.130855\\
15	0.088605\\
20	0.0665425\\
25	0.05963\\
30	0.0549325\\
};
\addlegendentry{\scriptsize{Model 1, Tech 1}}

\addplot [color=mycolor2, line width=2.0pt, mark=square, mark options={solid, mycolor2}]
  table[row sep=crcr]{%
5	0.19752125\\
10	0.12609925\\
15	0.08508825\\
20	0.06593625\\
25	0.05766025\\
30	0.054848\\
};
\addlegendentry{\scriptsize{Model 1, Tech 2}}

\addplot [color=mycolor3, line width=2.0pt, mark=x, mark options={solid, mycolor3}]
  table[row sep=crcr]{%
5	0.1892325\\
10	0.1218375\\
15	0.0764175\\
20	0.055645\\
25	0.051075\\
30	0.0469725\\
};
\addlegendentry{\scriptsize{Model 2, Tech 2, Gaussian}}

\addplot [color=mycolor4, dashdotted, line width=3.0pt, mark=triangle, mark options={solid, rotate=180, mycolor4}]
  table[row sep=crcr]{%
5	0.1873475\\
10	0.1199725\\
15	0.07259\\
20	0.0479425\\
25	0.039425\\
30	0.0348875\\
};
\addlegendentry{\scriptsize{Model 2, Tech 1, Gaussian}}

\addplot [color=mycolor5, line width=2.0pt, mark=o, mark options={solid, mycolor5}]
  table[row sep=crcr]{%
5	0.18471\\
10	0.119185\\
15	0.072965\\
20	0.04738\\
25	0.0383725\\
30	0.03324\\
};
\addlegendentry{\scriptsize{Model 2, Tech 1, Polynomial}}

\addplot [color=black, line width=1.0pt, mark=diamond, mark options={solid, black}]
  table[row sep=crcr]{%
5	0.18736125\\
10	0.11459975\\
15	0.0702405\\
20	0.0456965\\
25	0.034432\\
30	0.0298715\\
};
\addlegendentry{\scriptsize{Maximum Likelihood Decoder}}

\end{axis}
\end{tikzpicture}%
		\caption{Bit-error probability vs. signal-to-noise ratio (dB) when $\alpha = 0.95$, $0.97$ and $0.99$, respectively.}\label{fig:fig_7}
	\end{center}
	\vspace{-0.5cm}
\end{figure}
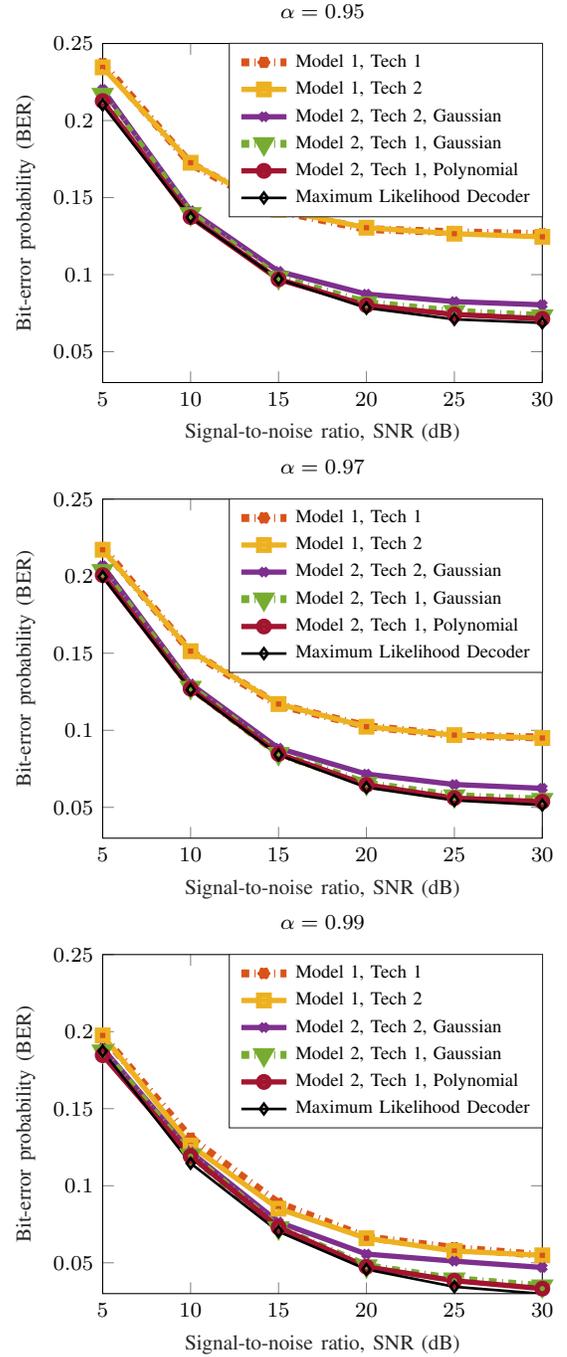
\emph{Technique 1}: There is only one \ac{SVM} classifier that outputs the class of encoding codeword entering the encoder at the transmitter. We display the receiver model with the technique in Fig. \ref{fig:fig_5}. Specifically, because we have $N$ bits in one encoding codeword, i.e., $\textbf{b}_{i}$ is $N\times1$, we have $2^{N}$ classes, which is the number of elements in set $\mathcal{B}$, and each class represents one encoding codeword, and hence one element of $\mathcal{B}$. Here, the classification is multi-nominal, and the classifier may perform either one-vs.-one or one-vs.-rest technique.

\emph{Technique 2}: We employ $N$ classifiers, where each classifier outputs the class of one element of $\textbf{b}_{i}$, i.e., either one or zero. As seen in Fig. \ref{fig:fig_4}, we have again $[\textbf{r}_{i+1}\;\textbf{r}_{i}\;\textbf{z}_{i}]$ as the input to the $N$ different classifiers. Because each element of the encoding codeword is either one or zero, we train one binary classifier for each element of the vector. The reason behind employing this technique is that we have at most $N$ binary classifiers, whereas we have one multi-class classifier with $2^{N}$ classes in the first technique, and the number of classes increases with increasing $N$. 

\section{Performance Analysis}\label{sec:performance_analysis}
We consider the aforementioned two receiver models, and compare their bit-error-rate performance with the maximum likelihood decoder. We consider that the transmitter employs linear block codes, particularly Hamming (7,4) codes. Specifically, we set $N = 4$ and $M = 7$ at the encoder. We use the binary-phase-shift keying (BPSK) method in the modulator and employ one pilot symbol in each frame; therefore, we have $L = 1$ and $T = 7$ at the modulator output, where the first symbol is the pilot symbol and the other $7$ symbols are the data-bearing BPSK symbols. Nevertheless, one can compare these two models having different linear block code and data modulation techniques with more than one pilot symbols. Furthermore, we define the signal-to-noise ratio as $\tSNR = \frac{P}{\sigma_{w}^{2}}$.

In Technique 1 of Model 2, we deploy the one-vs.-one technique. The SVM classifier compares all the classes in combinations of two for the given input vector and runs a \emph{max-win} voting strategy in order to finalize the classification. Specifically, the classifier assigns the instance to one class of all the class combinations of two, and acknowledges the class with the most votes as the winner. On the other hand, in Technique 2 of Model 2, the classifier assigns either $1$ or $0$ to each and every input at each classifier, which corresponds to one bit of $\textbf{b}_{i}$. We omit the polynomial function for Technique 2 in our figures because we observe in our simulations that Technique 2 with the Gaussian radial basis function outperforms Technique 2 with the polynomial function significantly. Herein, for the sake of simplicity and clarity in the presentation of the paper, we omit the details of SVMs, and refer interested readers to \cite{thompson2019stat} for more details since we use the machine learning toolbox from MATLAB.

\begin{figure}[t]
	\begin{center}
%
%
\definecolor{mycolor1}{rgb}{0.63500,0.07800,0.18400}%
\definecolor{mycolor2}{rgb}{0.85000,0.32500,0.09800}%
\begin{tikzpicture}

\begin{axis}[%
font = \footnotesize,
width=4.602in*\scaleFigure,
height=3.552in*\scaleFigure,
at={(0.772in*\scaleFigure,0.495in*\scaleFigure)},
scale only axis,
xmin=5,
xmax=30,
xlabel style={font=\color{white!15!black}\footnotesize},
xlabel={Signal-to-noise ratio, SNR (dB)},
ymin=0.03,
ymax=0.25,
ylabel style={font=\color{white!15!black}\footnotesize},
ylabel={Bit-error probability (BER)},
axis background/.style={fill=white},
title style={font=\footnotesize},
legend style={at={(0.283*\scaleFigure,1.2925*\scaleFigure)}, anchor=south west, legend cell align=left, align=left, draw=white!15!black}
]
\addplot [color=mycolor1, dashdotted, line width=2.0pt, mark=o, mark options={solid, mycolor1}]
  table[row sep=crcr]{%
5	0.246745\\
10	0.16657\\
15	0.12657\\
20	0.105465\\
25	0.099005\\
30	0.095855\\
};
\addlegendentry{\scriptsize{Model 2, Tech 1, Polynomial (1-Bit)}}

\addplot [color=mycolor2, dashdotted, line width=2.0pt, mark=asterisk, mark options={solid, mycolor2}]
  table[row sep=crcr]{%
5	0.2185225\\
10	0.15103\\
15	0.1164875\\
20	0.102695\\
25	0.096585\\
30	0.0953375\\
};
\addlegendentry{\scriptsize{Model 1, Tech 1, (32-Bit)}}

\addplot [color=mycolor1, line width=2.0pt, mark=o, mark options={solid, mycolor1}]
  table[row sep=crcr]{%
5	0.2007\\
10	0.126975\\
15	0.08468\\
20	0.06462\\
25	0.056\\
30	0.05365\\
};
\addlegendentry{\scriptsize{Model 2, Tech 1 Polynomial (32-Bit)}}

\addplot [color=black, line width=2.0pt, mark=diamond, mark options={solid, black}]
  table[row sep=crcr]{%
5	0.19966925\\
10	0.1262915\\
15	0.08404625\\
20	0.06288375\\
25	0.05453175\\
30	0.05148425\\
};
\addlegendentry{\scriptsize{Maximum Likelihood Decoder (32-Bit)}}

\end{axis}
\end{tikzpicture}%
\\
%
%
\definecolor{mycolor1}{rgb}{0.46600,0.67400,0.18800}%
\definecolor{mycolor2}{rgb}{0.85000,0.32500,0.09800}%
\begin{tikzpicture}

\begin{axis}[%
font = \footnotesize,
width=4.602in*\scaleFigure,
height=3.552in*\scaleFigure,
at={(0.772in*\scaleFigure,0.495in*\scaleFigure)},
scale only axis,
xmin=5,
xmax=30,
xlabel style={font=\color{white!15!black}\footnotesize},
xlabel={Signal-to-noise ratio, SNR (dB)},
ymin=0.03,
ymax=0.25,
ylabel style={font=\color{white!15!black}\footnotesize},
ylabel={Bit-error probability (BER)},
axis background/.style={fill=white},
title style={font=\footnotesize},
legend style={at={(0.283*\scaleFigure,1.2925*\scaleFigure)}, anchor=south west, legend cell align=left, align=left, draw=white!15!black}
]
\addplot [color=mycolor1, dashdotted, line width=2.0pt, mark=triangle, mark options={solid, rotate=180, mycolor1}]
  table[row sep=crcr]{%
5	0.246225\\
10	0.1674825\\
15	0.127515\\
20	0.1060575\\
25	0.0991775\\
30	0.0967\\
};
\addlegendentry{\scriptsize{Model 2, Tech 1, Gaussian (1-Bit)}}

\addplot [color=mycolor2, dashdotted, line width=2.0pt, mark=asterisk, mark options={solid, mycolor2}]
  table[row sep=crcr]{%
5	0.2185225\\
10	0.15103\\
15	0.1164875\\
20	0.102695\\
25	0.096585\\
30	0.0953375\\
};
\addlegendentry{\scriptsize{Model 1, Tech 1, (32-Bit)}}

\addplot [color=mycolor1, line width=2.0pt, mark=triangle, mark options={solid, rotate=180, mycolor1}]
  table[row sep=crcr]{%
5	0.2035425\\
10	0.127815\\
15	0.08493\\
20	0.0654775\\
25	0.0573775\\
30	0.05504\\
};
\addlegendentry{\scriptsize{Model 2, Tech 1, Gaussian (32-Bit)}}

\addplot [color=black, line width=2.0pt, mark=diamond, mark options={solid, black}]
  table[row sep=crcr]{%
5	0.19966925\\
10	0.1262915\\
15	0.08404625\\
20	0.06288375\\
25	0.05453175\\
30	0.05148425\\
};
\addlegendentry{\scriptsize{Maximum Likelihood Decoder (32-Bit)}}

\end{axis}
\end{tikzpicture}%
\\
%
%
\definecolor{mycolor1}{rgb}{0.49400,0.18400,0.55600}%
\definecolor{mycolor2}{rgb}{0.85000,0.32500,0.09800}%
\begin{tikzpicture}

\begin{axis}[%
font = \footnotesize,
width=4.602in*\scaleFigure,
height=3.552in*\scaleFigure,
at={(0.772in*\scaleFigure,0.495in*\scaleFigure)},
scale only axis,
xmin=5,
xmax=30,
xlabel style={font=\color{white!15!black}\footnotesize},
xlabel={Signal-to-noise ratio, SNR (dB)},
ymin=0.03,
ymax=0.25,
ylabel style={font=\color{white!15!black}\footnotesize},
ylabel={Bit-error probability (BER)},
axis background/.style={fill=white},
title style={font=\footnotesize},
legend style={at={(0.283*\scaleFigure,1.2925*\scaleFigure)}, anchor=south west, legend cell align=left, align=left, draw=white!15!black}
]
\addplot [color=mycolor1, dashdotted, line width=2.0pt, mark=x, mark options={solid, mycolor1}]
  table[row sep=crcr]{%
5	0.244885\\
10	0.1660025\\
15	0.1255075\\
20	0.1050475\\
25	0.09873\\
30	0.0961175\\
};
\addlegendentry{\scriptsize{Model 2, Tech 2, Gaussian (1-Bit)}}

\addplot [color=mycolor2, dashdotted, line width=2.0pt, mark=asterisk, mark options={solid, mycolor2}]
  table[row sep=crcr]{%
5	0.2185225\\
10	0.15103\\
15	0.1164875\\
20	0.102695\\
25	0.096585\\
30	0.0953375\\
};
\addlegendentry{\scriptsize{Model 1, Tech 1, (32-Bit)}}

\addplot [color=mycolor1, line width=2.0pt, mark=x, mark options={solid, mycolor1}]
  table[row sep=crcr]{%
5	0.2067675\\
10	0.1300675\\
15	0.08838\\
20	0.071595\\
25	0.0648325\\
30	0.0623475\\
};
\addlegendentry{\scriptsize{Model 2, Tech 2, Gaussian (32-Bit)}}

\addplot [color=black, line width=2.0pt, mark=diamond, mark options={solid, black}]
  table[row sep=crcr]{%
5	0.19966925\\
10	0.1262915\\
15	0.08404625\\
20	0.06288375\\
25	0.05453175\\
30	0.05148425\\
};
\addlegendentry{\scriptsize{Maximum Likelihood Decoder (32-Bit)}}

\end{axis}

\end{tikzpicture}%
		\caption{Bit-error probability vs. signal-to-noise ratio (dB) when $\alpha = 0.97$. The top figure: Model 2 Technique 1 with the polynomial kernel. The middle figure: Model 2 Technique 1 with the Gaussian kernel. The bottom figure: Model 2 Technique 2 with the Gaussian kernel.}\label{fig:fig_8}
	\end{center}
	\vspace{-0.5cm}
\end{figure}
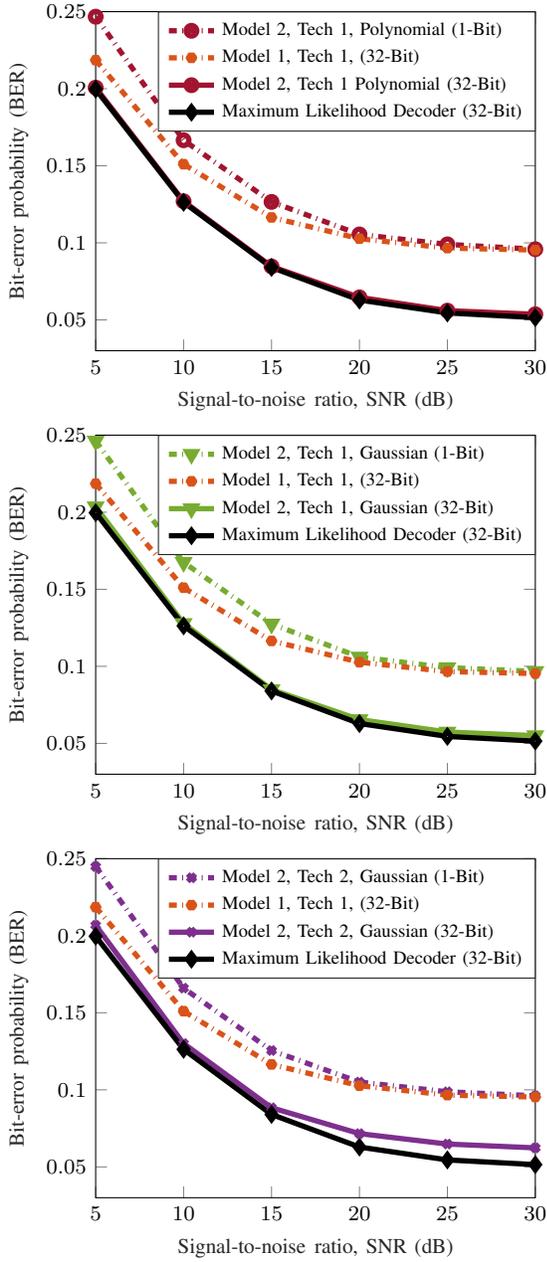

In Fig. \ref{fig:fig_7}, we plot the bit-error probability as a function of the signal-to-noise ratio. We set $\alpha = 0.95$, $0.97$ and $0.99$, respectively. In all the cases of $\alpha$, Model 2 Technique 1 with the degree-2 polynomial and auto-scaled Gaussian radial basis kernels perform very close the maximum likelihood decoder, which is the optimum decoder when the encoder inputs are equally likely at the transmitter. When compared to Model 1 Techniques 1 and 2, where MMSE and Kalman filter estimators are implemented in the channel estimation process, respectively, Model 2 with the given SVM parameters has lower bit-error probability, and the performance gap between Model 1 and Model 2 increases with decreasing $\alpha$, i.e., the decreasing correlation between the channel fading gains over time. Furthermore, Model 2 Technique 2, i.e., the receiver model with $N$ different classifiers, also performs very closely to the maximum likelihood detector when the Gaussian radial basis kernel is employed. Considering the results in Fig. \ref{fig:fig_7}, one can see that we can increase the symbol detection performance in fast fading channels by running SVM-based techniques. Moreover, we can avoid decoding complexity by implementing a receiver with multiple SVM classifiers as in the case of Model 2 Technique 2.
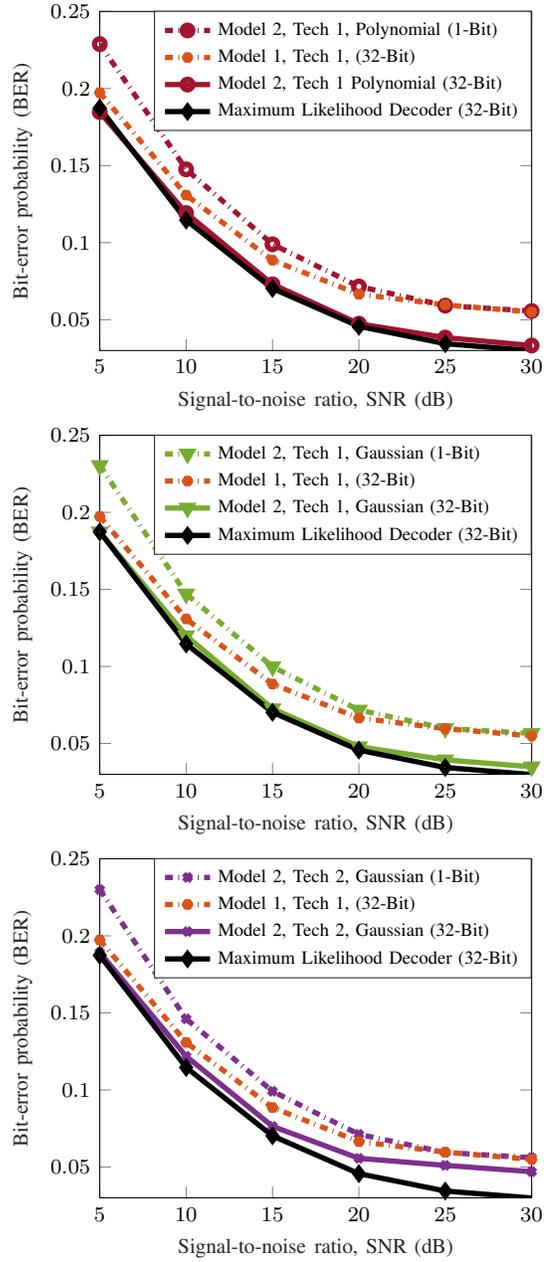
\begin{figure}[t]
	\begin{center}
%
%
\definecolor{mycolor1}{rgb}{0.63500,0.07800,0.18400}%
\definecolor{mycolor2}{rgb}{0.85000,0.32500,0.09800}%
\begin{tikzpicture}

\begin{axis}[%
font = \footnotesize,
width=4.521in*\scaleFigure,
height=3.552in*\scaleFigure,
at={(0.758in*\scaleFigure,0.495in*\scaleFigure)},
scale only axis,
xmin=5,
xmax=30,
xlabel style={font=\color{white!15!black}\footnotesize},
xlabel={Signal-to-noise ratio, SNR (dB)},
ymin=0.03,
ymax=0.25,
ylabel style={font=\color{white!15!black}\footnotesize},
ylabel={Bit-error probability (BER)},
axis background/.style={fill=white},
title style={font=\footnotesize},
legend style={at={(0.251*\scaleFigure,1.295*\scaleFigure)}, anchor=south west, legend cell align=left, align=left, draw=white!15!black}
]
\addplot [color=mycolor1, dashdotted, line width=2.0pt, mark=o, mark options={solid, mycolor1}]
  table[row sep=crcr]{%
5	0.22884\\
10	0.14755\\
15	0.0989325\\
20	0.07155\\
25	0.0591325\\
30	0.055605\\
};
\addlegendentry{\scriptsize{Model 2, Tech 1, Polynomial (1-Bit)}}

\addplot [color=mycolor2, dashdotted, line width=2.0pt, mark=asterisk, mark options={solid, mycolor2}]
  table[row sep=crcr]{%
5	0.197325\\
10	0.130855\\
15	0.088605\\
20	0.0665425\\
25	0.05963\\
30	0.0549325\\
};
\addlegendentry{\scriptsize{Model 1, Tech 1, (32-Bit)}}

\addplot [color=mycolor1, line width=2.0pt, mark=o, mark options={solid, mycolor1}]
  table[row sep=crcr]{%
5	0.18471\\
10	0.119185\\
15	0.072965\\
20	0.04738\\
25	0.0383725\\
30	0.03324\\
};
\addlegendentry{\scriptsize{Model 2, Tech 1 Polynomial (32-Bit)}}

\addplot [color=black, line width=2.0pt, mark=diamond, mark options={solid, black}]
  table[row sep=crcr]{%
5	0.18736125\\
10	0.11459975\\
15	0.0702405\\
20	0.0456965\\
25	0.034432\\
30	0.0298715\\
};
\addlegendentry{\scriptsize{Maximum Likelihood Decoder (32-Bit)}}

\end{axis}
\end{tikzpicture}%
\\
%
%
\definecolor{mycolor1}{rgb}{0.46600,0.67400,0.18800}%
\definecolor{mycolor2}{rgb}{0.85000,0.32500,0.09800}%
\begin{tikzpicture}

\begin{axis}[%
font = \footnotesize,
width=4.521in*\scaleFigure,
height=3.552in*\scaleFigure,
at={(0.758in*\scaleFigure,0.495in*\scaleFigure)},
scale only axis,
xmin=5,
xmax=30,
xlabel style={font=\color{white!15!black}\footnotesize},
xlabel={Signal-to-noise ratio, SNR (dB)},
ymin=0.03,
ymax=0.25,
ylabel style={font=\color{white!15!black}\footnotesize},
ylabel={Bit-error probability (BER)},
axis background/.style={fill=white},
title style={font=\footnotesize},
legend style={at={(0.251*\scaleFigure,1.295*\scaleFigure)}, anchor=south west, legend cell align=left, align=left, draw=white!15!black}
]
\addplot [color=mycolor1, dashdotted, line width=2.0pt, mark=triangle, mark options={solid, rotate=180, mycolor1}]
  table[row sep=crcr]{%
5	0.230425\\
10	0.14706\\
15	0.0998525\\
20	0.0717325\\
25	0.0598025\\
30	0.056545\\
};
\addlegendentry{\scriptsize{Model 2, Tech 1, Gaussian (1-Bit)}}

\addplot [color=mycolor2, dashdotted, line width=2.0pt, mark=asterisk, mark options={solid, mycolor2}]
  table[row sep=crcr]{%
5	0.197325\\
10	0.130855\\
15	0.088605\\
20	0.0665425\\
25	0.05963\\
30	0.0549325\\
};
\addlegendentry{\scriptsize{Model 1, Tech 1, (32-Bit)}}

\addplot [color=mycolor1, line width=2.0pt, mark=triangle, mark options={solid, rotate=180, mycolor1}]
  table[row sep=crcr]{%
5	0.1873475\\
10	0.1199725\\
15	0.07259\\
20	0.0479425\\
25	0.039425\\
30	0.0348875\\
};
\addlegendentry{\scriptsize{Model 2, Tech 1, Gaussian (32-Bit)}}

\addplot [color=black, line width=2.0pt, mark=diamond, mark options={solid, black}]
  table[row sep=crcr]{%
5	0.18736125\\
10	0.11459975\\
15	0.0702405\\
20	0.0456965\\
25	0.034432\\
30	0.0298715\\
};
\addlegendentry{\scriptsize{Maximum Likelihood Decoder (32-Bit)}}

\end{axis}
\end{tikzpicture}%
\\
%
%
\definecolor{mycolor1}{rgb}{0.49400,0.18400,0.55600}%
\definecolor{mycolor2}{rgb}{0.85000,0.32500,0.09800}%
\begin{tikzpicture}

\begin{axis}[%
font = \footnotesize,
width=4.521in*\scaleFigure,
height=3.552in*\scaleFigure,
at={(0.758in*\scaleFigure,0.495in*\scaleFigure)},
scale only axis,
xmin=5,
xmax=30,
xlabel style={font=\color{white!15!black}\footnotesize},
xlabel={Signal-to-noise ratio, SNR (dB)},
ymin=0.03,
ymax=0.25,
ylabel style={font=\color{white!15!black}\footnotesize},
ylabel={Bit-error probability (BER)},
axis background/.style={fill=white},
title style={font=\footnotesize},
legend style={at={(0.251*\scaleFigure,1.295*\scaleFigure)}, anchor=south west, legend cell align=left, align=left, draw=white!15!black}
]
\addplot [color=mycolor1, dashdotted, line width=2.0pt, mark=x, mark options={solid, mycolor1}]
  table[row sep=crcr]{%
5	0.229995\\
10	0.146135\\
15	0.0991475\\
20	0.071215\\
25	0.059435\\
30	0.056065\\
};
\addlegendentry{\scriptsize{Model 2, Tech 2, Gaussian (1-Bit)}}

\addplot [color=mycolor2, dashdotted, line width=2.0pt, mark=asterisk, mark options={solid, mycolor2}]
  table[row sep=crcr]{%
5	0.197325\\
10	0.130855\\
15	0.088605\\
20	0.0665425\\
25	0.05963\\
30	0.0549325\\
};
\addlegendentry{\scriptsize{Model 1, Tech 1, (32-Bit)}}

\addplot [color=mycolor1, line width=2.0pt, mark=x, mark options={solid, mycolor1}]
  table[row sep=crcr]{%
5	0.1892325\\
10	0.1218375\\
15	0.0764175\\
20	0.055645\\
25	0.051075\\
30	0.0469725\\
};
\addlegendentry{\scriptsize{Model 2, Tech 2, Gaussian (32-Bit)}}

\addplot [color=black, line width=2.0pt, mark=diamond, mark options={solid, black}]
  table[row sep=crcr]{%
5	0.18736125\\
10	0.11459975\\
15	0.0702405\\
20	0.0456965\\
25	0.034432\\
30	0.0298715\\
};
\addlegendentry{\scriptsize{Maximum Likelihood Decoder (32-Bit)}}

\end{axis}
\end{tikzpicture}%
		\caption{Bit-error probability vs. signal-to-noise ratio (dB) when $\alpha = 0.99$. The top figure: Model 2 Technique 1 with the polynomial kernel. The middle figure: Model 2 Technique 1 with the Gaussian kernel. The bottom figure: Model 2 Technique 2 with the Gaussian kernel.}\label{fig:fig_9}
	\end{center}
	\vspace{-0.5cm}
\end{figure}

\subsection{1-Bit Analog-to-Digital Converter}
Low-resolution \acp{ADC} have recently attracted much attention in wireless communication settings that require low-power consumption and less computation cost. For example, massive \ac{MIMO} \cite{li2017channel,8487039} and \ac{OFDM} \cite{balevi2019one} receivers have become the reasons to implement 1-bit \acp{ADC}. We further know that in supervised machine learning-based applications, after the training process, memory may still be a concern in the prediction process in addition to low-power consumption and less computation requirements. As seen in Fig. \ref{fig:fig_7}, the SVM-based receiver performs very well in fast fading channels when compared to the conventional receivers employing channel estimation, demodulation and data decoding as separate blocks. Moreover, we know that SVMs perform addition of a weighted combination of support vectors during prediction, and these support vectors should be stored after an SVM-based receiver is trained. If there are a lot of support vectors, it may be a problem if we want to deploy our SVM-based receiver in low-memory devices, e.g., mobile phones and wireless sensors. However, we can attack this problem by either decreasing the number of support vectors or implementing 1-Bit \acp{ADC}, which leads to a certain performance and memory tradeoff. Therefore, we update the receiver models given in Fig. \ref{fig:fig_5} and Fig. \ref{fig:fig_6} such that the input to the classifiers become the signs of the elements of $\textbf{z}_{i}$, $\textbf{r}_{i}$ and $\textbf{r}_{i+1}$.

In Fig. \ref{fig:fig_8} and Fig. \ref{fig:fig_9}, we plot the 1-bit SVM-based decoder results along with their corresponding 32-bit \ac{SVM}-based decoder results and the 32-bit Model 1 Technique 1 and maximum likelihood decoder results when $\alpha = 0.97$ and $0.99$, respectively. The 1-bit \ac{SVM}-based decoder performs very close to 32-bit Model 1 Technique 1 in signal-to-noise ratio regimes above 20 dB in all the cases of $\alpha$. Here, we compare the 1-bit \ac{SVM}-based decoder results with the 32-bit results in order to show the tradeoff between the computational complexity and cost, and the performance loss due to a simplified design. Hence, one can resort to 1-bit \ac{SVM} depending on the type and quality-of-service constraints of transmitted data and the channel conditions for a certain amount of performance loss in order to gain design advantages.
~
\section{Conclusion}\label{sec:conclusion}
In this paper, we have investigated the bit-error probability performance of \ac{SVM}-based receivers considering different techniques and 1-bit \acp{ADC} in fast and flat fading channels. We propose two different techniques for the \ac{SVM}-based receivers that perform channel estimation, data demodulation and decoding jointly. One technique is to classify the received signal into one class representing the decimal value of the encoding codeword at the input of the encoder at the transmitter side. The other technique is to use one binary classifier for each bit of the encoding codeword. We have compared the bit-error probability of these techniques with the one of the maximum likelihood detector, which is know to be the optimum detector when the encoding codewords are equally likely. We have shown that the \ac{SVM}-based receiver can perform very close to the optimal values. We have also shown that one can simplify the \ac{SVM}-based receiver design and still reach reasonable performance levels. This work also shows that we can deal with the increasing complexity in future technologies and reach the optimal bit error probabilities by invoking machine learning tools in receiver design.
 
\bibliographystyle{IEEEtran}
\bibliography{references}

\end{document}